\documentclass[twocolumn,showpacs,prb,amsmath,amssymb,floatfix,eqsecnum]{revtex4-1}

\usepackage{amsmath,amssymb,color}
\usepackage{bm}
\usepackage{epsfig}
\usepackage{psfrag}
\usepackage{hyperref}

\newcommand{\bd}{\bm}

\begin{document}

\title{Magneto-elastic modes and lifetime of magnons
in thin yttrium-iron garnet films}

\author{Andreas R\"{u}ckriegel}
\affiliation{Institut f\"{u}r Theoretische Physik, Universit\"{a}t Frankfurt,  Max-von-Laue Strasse 1, 60438 Frankfurt, Germany}
\affiliation{Department of Physics, University of Florida, Gainesville, 
Florida 32611, USA}

\author{Peter Kopietz}
\affiliation{Institut f\"{u}r Theoretische Physik, Universit\"{a}t Frankfurt,  Max-von-Laue Strasse 1, 60438 Frankfurt, Germany}
\affiliation{Department of Physics, University of Florida, Gainesville, 
Florida 32611, USA}

\author{Dmytro A. Bozhko}
\affiliation{Fachbereich Physik and Landesforschungszentrum OPTIMAS, Technische Universit\"at Kaiserslautern, 67663 Kaiserslautern, Germany}

\author{Alexander A. Serga}
\affiliation{Fachbereich Physik and Landesforschungszentrum OPTIMAS, Technische Universit\"at Kaiserslautern, 67663 Kaiserslautern, Germany}

\author{Burkard Hillebrands}
\affiliation{Fachbereich Physik and Landesforschungszentrum OPTIMAS, Technische Universit\"at Kaiserslautern, 67663 Kaiserslautern, Germany}

\date{May 20, 2014}
%\date{\today}

 \begin{abstract}
We calculate the effects of the spin-lattice coupling
on the magnon spectrum of
thin ferromagnetic films consisting of the magnetic insulator
yttrium-iron garnet. 
The magnon-phonon hybridisation generates
a characteristic minimum in
the spin dynamic structure factor which quantitatively agrees with
recent Brillouin light scattering experiments.  
We also show that at room temperature
the phonon contribution to the  magnon damping 
exhibits a rather complicated momentum 
dependence:
In the exchange regime the magnon damping is dominated by
Cherenkov type scattering processes,
while in the long-wavelength dipolar regime
these processes are subdominant and the
magnon damping is two orders of magnitude smaller.
We supplement our calculations by actual measurements
of the magnon relaxation
in the dipolar regime.
Our theory provides a simple explanation of a recent experiment probing the
different temperatures of the magnon and phonon gases in yttrium-iron garnet.
\end{abstract}


\maketitle

\section{Introduction}

Although the spin-lattice interactions in magnetic insulators 
can often be ignored, in some cases the coupling between the spin  degrees 
of freedom and the lattice vibrations (phonons) plays a crucial role.
For example, in ultrasound experiments
one uses the spin-lattice coupling to study
the properties of the spin degrees of freedom from the 
measurement of the propagation and the attenuation of sound 
waves~\cite{Luethi05}.
%But also for the proper interpretation of
%Brillouin light scattering experiments in
%magnetic insulators
%spin-lattice interactions can play an important role.
The theory of
magneto-elastic effects
in magnetic insulators has been developed more than 
half a century ago by Abrahams and Kittel \cite{Abrahams52,Kittel58}, and by
Kaganov and Tsukernik \cite{Kaganov59}. 
While in the past decades a few theoretical studies of
magneto-elastic effects have appeared 
\cite{Tiersten64,Bandari66,Lord68,Kobayashi73,Gurevich96,Dreher12},
recent experimental progress in the field
of spintronics has revived the interest in the interactions
between spin and lattice degrees of freedom~\cite{Kamra13}.

The present work is motivated by
an experiment \cite{Agrawal13} where the coupling between
magnons and phonons
in the magnetic insulator yttrium iron garnet (YIG) was 
probed via a spatially 
resolved measurement of the magnon temperature $T_m$
in the presence of a thermal gradient. In the 
short wavelength exchange part of the magnon spectrum 
$T_m$ was found to agree almost perfectly with the 
temperature $T_p$ of the phonon bath. 
In order to reconcile
this finding with earlier studies of the spin Seebeck effect \cite{Uchida08}
(which relies on the difference between $T_m$ and $T_p$),
the authors of Ref.~[\onlinecite{Agrawal13}] speculated
that in the long-wavelength 
dipolar part of the spectrum the magnon temperature significantly differs from $T_p$, suggesting a rather weak coupling between magnons and phonons 
in this regime.
In this work we offer a microscopic explanation for such a
momentum dependence of the magnon temperature: We show that the lifetime
$\tau ( \bd{k} )$
of magnons due to coupling to the phonons in YIG 
is strongly momentum dependent;
 in particular $\tau ( \bd{k} )$
exhibits a pronounced minimum in the exchange regime
and is two orders of magnitude larger in the
dipolar regime.
Since in the dipolar part of the spectrum
the magnons have a longer lifetime, in this regime 
phonons are less effective to
thermalize the magnons and the differences between $T_m$ and $T_p$
can persist for longer times.

We also present experimental results
for the magnon damping in the dipolar regime, which have been
obtained by means of
time- and wave-vector-resolved Brillouin light scattering (BLS) 
spectroscopy \cite{Sandweg}.
The experimentally determined damping rate is three orders of magnitude larger
than our theoretical prediction, although the qualitative behavior
as a function of in-plane momentum is similar.
One should keep in mind, however,
that in our calculations only the relaxation due to
magnon-phonon interactions has been taken into account. Apparently, in the
long-wavelength dipolar regime other scattering channels leading to momentum relaxation are dominant,
such as elastic magnon-impurity scattering.

The rest of this work is organized as follows: In Sec.~\ref{sec:magnons}
we briefly review the calculation of the spin wave spectrum
of a thin YIG stripe and fix the experimentally relevant parameters.
In Sec.~\ref{sec:magpho} we carefully derive the
magnon-phonon interaction by quantizing the
phenomenological classical magneto-elastic energy. 
The calculation of the magneto-elastic modes  due to the magnon-phonon
hybridisation is presented in Sec.~\ref{sec:hybrid}. 
We also calculate the resulting spectral function of the magnons and
the transverse dynamic structure factor which is proportional
to the BLS cross section.
Since we are
interested in the magnon dynamics, we derive the effective action of the
magnons by  integrating over the phonon degrees of freedom.
Using this effective action, we 
proceed in Sec.~\ref{sec:damp} to calculate the damping of the magnons
due to the coupling to the phonons. We also present new 
experimental
results for the magnon damping in the long-wavelength dipolar regime,
and compare them to our calculations.
Finally, in Sec.~\ref{sec:conclusions} we present our conclusions.
To make contact with previous work \cite{Gurevich96} on magnon-phonon interactions in YIG, we present in the appendix an alternative derivation
of the dispersion of the magneto-elastic modes using the equations of motion.

\section{Magnons in YIG}
\label{sec:magnons}

\begin{figure}[t]
\includegraphics[width=70mm]{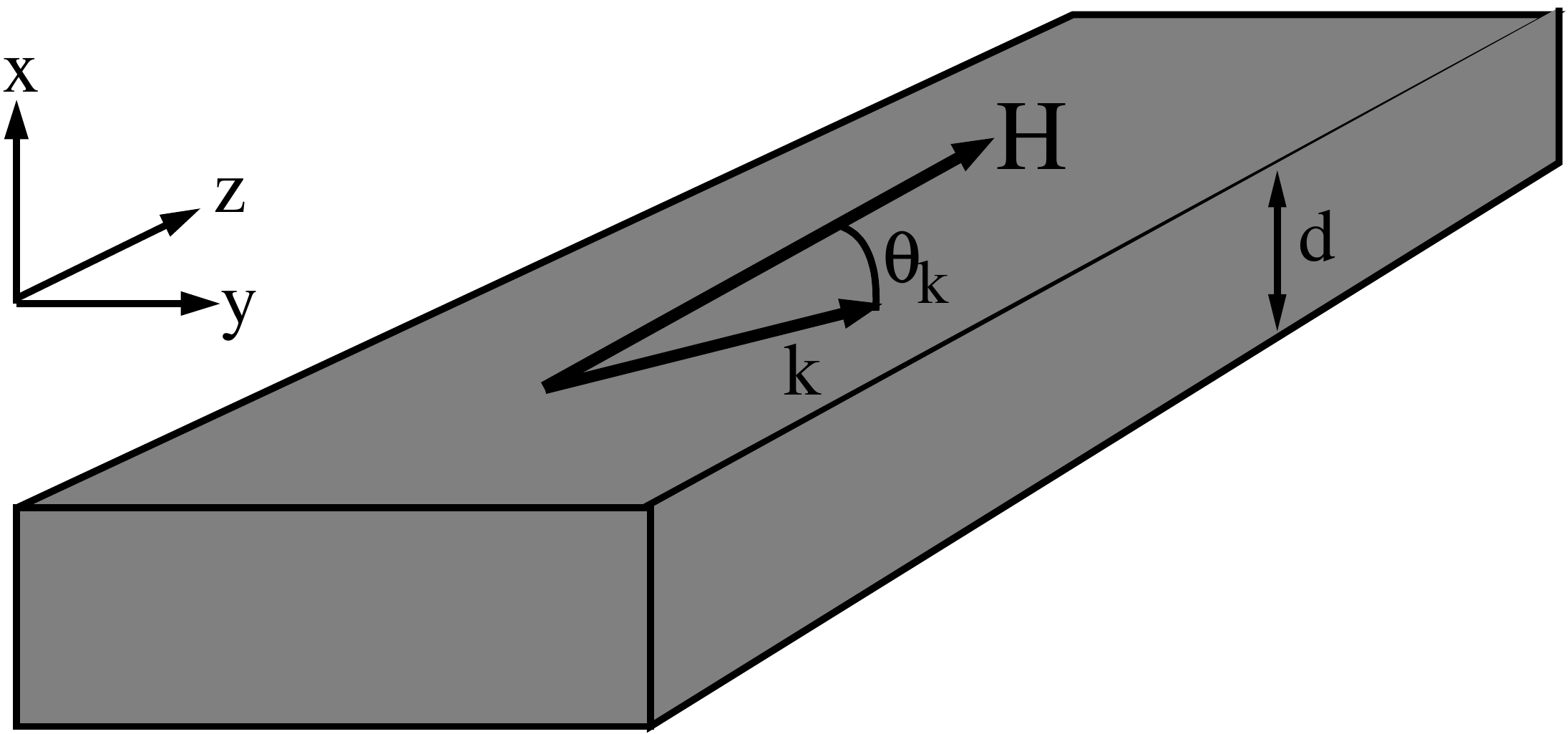}
  \caption{%
Thin YIG stripe of thickness $d$ in a magnetic
field $\bd{H} = H \bd{e}_z$ along the direction
$\bd{e}_z$ of the long axis.
We consider magnons with wave vector
$\bd{k} = | \bd{k} |  \cos \theta_{\bd{k}} \bd{e}_z +
 | \bd{k} | \sin \theta_{\bd{k}} \bd{e}_y$ in the stripe-plane.
}
  \label{fig:geometry}
\end{figure}

It is generally established that
the magnetic properties of YIG at room temperature
can be obtained from the following effective quantum spin 
Hamiltonian \cite{Cherepanov93,Tupitsyn08,Kreisel09} 
 \begin{eqnarray} 
 {\cal{H}}  &=& 
% -\frac{1}{2} \sum_{ij}  J_{ij} {\bd{S}}_i \cdot {\bd{S}}_j 
   - \frac{1}{2} \sum_{ij}
 \sum_{\alpha \beta}  ( J_{ij} \delta_{\alpha \beta} + 
 D_{ij}^{\alpha \beta} ) S_i^\alpha S_j^\beta
 - h \sum_i S^z_i ,
 \hspace{7mm}
 \label{eq:Hspin}
\end{eqnarray} 
where the  spin operators $\bd{S}_i = \bd{S} ( \bd{R}_i )$
are localized at the
sites $\bd{R}_i$ of a cubic lattice with
lattice spacing $a \approx 12.376 \, {\rm \AA}$, 
the exchange couplings $J_{ij} = J ( \bd{R}_i - \bd{R}_j )$ connect
the spins at nearest neighbor sites $\bd{R}_i$ and $\bd{R}_j$, and
$h = \mu H$ is the Zeemann energy due to an 
external magnetic field $H$ along the $z$ axis
(where $\mu = 2 \mu_B$, and $\mu_B$ is the Bohr magneton). 
The dipole-dipole interaction is
 \begin{equation}
 D_{ij}^{\alpha \beta}  = (1- \delta_{ij})\frac{\mu^2}{|\bd{R}_{ij}|^3} \left[ 3\hat R_{ij}^\alpha \hat R_{ij}^\beta - \delta_{\alpha \beta}  \right],
\end{equation}  
where ${\bd R}_{ij}={\bd R}_i - {\bd R}_j$ and $\hat {\bd R}_{ij} = 
{\bd R}_{ij}/|{\bd R}_{ij}|$.
Within the framework of the usual expansion
in inverse powers of the spin $S$  
the low-energy magnon spectrum of YIG can be
quantitatively described if one chooses
$J \approx 3.19 \, \mbox{\rm K}$ and $S = M_s a^3 / \mu \approx 14.2$,
where the saturation magnetization of YIG is given by
$4 \pi M_s = 1750 \, \mbox{\rm G}$.
Due to the large value of the effective spin $S$
we may use the Holstein-Primakoff
transformation \cite{Holstein40} to express the spin
operators in terms of canonical boson operators
$b_i$ and $b_i^{\dagger}$ and
expand the square roots,
\begin{subequations}
\begin{eqnarray}
S_i^+ &=&  \sqrt{2S} \sqrt{1  - \frac{b^{\dagger}_i b_i}{2S} } \; b_i 
 \approx  \sqrt{2S} \left[  b_i - \frac{b_i^\dagger b_i b_i}{4S} \right],
 \label{eq:hp1}
 \hspace{7mm}
\\
S_i^- &= & 
 \sqrt{2S} 	b_i^{\dagger} \sqrt{1  - \frac{b^{\dagger}_i b_i}{2S} } 
 \approx \sqrt{2S} \left[  b_i^\dagger - \frac{b_i^\dagger b_i^\dagger b_i}{4S}  \right],  
 \label{eq:hp2}
\\
 S_i^z &=& S-b_i^\dagger b_i, 
 \label{eq:hp3}
 \end{eqnarray}
 \end{subequations}
where 
$S_i^+ = S_i^x + i S_i^y$ and  $S_i^- = S_i^x - i S_i^y$, and 
we have assumed that the magnetization of the systen is uniform
and points along the $z$-axis.
To describe a thin stripe
we can work with an effective
two-dimensional model, as explained in Ref.~[\onlinecite{Kreisel09}].
In the geometry shown in
Fig.~\ref{fig:geometry} the in-plane magnon wave vectors
are then of the form $\bd{k} = k_y \bd{e}_y + k_z \bd{e}_z
=  | \bd{k} |  \cos \theta_{\bd{k}} \bd{e}_z +
 | \bd{k} | \sin \theta_{\bd{k}} \bd{e}_y$.
Defining
 \begin{equation}
 b_{\bd{k}} = \frac{1}{\sqrt{N}} \sum_i e^{  -i \bd{k} \cdot {\bd{R}}_i } 
 b_{i},
 \end{equation}
where $N$ denotes the number of lattice sites in the $yz$-plane,
and retaining only quadratic terms in the bosons,
we obtain the bosonized Hamiltonian for YIG,
 \begin{equation}
  {\cal{H}}^{(2)}_{\rm m} = \sum_{\bd{k}}  
 \left[ A_{\bd{k}} b^{\dagger}_{\bd{k}} b_{\bd{k}}
 + \frac{B_{\bd{k}}}{2} 
 \left( b^{\dagger}_{\bd{k}} b^{\dagger}_{- \bd{k}} 
 + b_{- \bd{k}} b_{\bd{k}} \right) \right].
 \label{eq:Hm}
 \end{equation}
Here the energies $A_{\bd{k}}$ and $B_{\bd{k}}$ can be 
expressed in terms of the 
Fourier transforms $J_{\bd{k}}$
and $D^{\alpha \beta}_{\bd{k}}$
of the exchange couplings and the dipole-dipole interaction
as \cite{Kreisel09}
 \begin{subequations}
 \begin{eqnarray}
 A_{\bd{k}} & = & h +S ( J_0 - J_{\bd{k}} ) + S \left( D_0^{zz} -   
 \frac{ D_{\bd{k}}^{xx} + D_{\bd{k}}^{yy}}{2} \right),
 \hspace{7mm}
 \\
 B_{\bd{k}} & = &  -S 
 \frac{ D_{\bd{k}}^{xx} - D_{\bd{k}}^{yy}}{2}  .
 \end{eqnarray} 
 \end{subequations}
The Hamiltonian (\ref{eq:Hm}) is easily diagonalized by means
of a Bogoliubov transformation,
 \begin{equation}
 \left( \begin{array}{c} b_{\bd{k}} \\ b^{\dagger}_{ - \bd{k}} 
 \end{array} \right) =
 \left( \begin{array}{cc} u_{\bd{k}} & - v_{\bd{k}} \\
 - v_{\bd{k}}^{\ast} & u_{\bd{k}} \end{array} \right)
  \left( \begin{array}{c} \beta_{\bd{k}} \\ \beta^{\dagger}_{ - \bd{k}} 
  \end{array} \right),
 \label{eq:Bogo}
 \end{equation}
with
 \begin{equation}
 u_{\bd{k}} = \sqrt{ \frac{ A_{\bd{k}} + E_{\bd{k}} }{2 E_{\bd{k}} } },
 \; \; \; v_{\bd{k}} = \frac{ B_{\bd{k}}}{ | B_{\bd{k}} | }
 \sqrt{ \frac{ A_{\bd{k}} - E_{\bd{k}} }{2 E_{\bd{k}} } },
 \end{equation}
and
 \begin{equation}
 E_{\bd{k}} = \sqrt{ A_{\bd{k}}^2 - B_{\bd{k}}^2 }.
 \label{eq:Ekdef}
 \end{equation}
In terms of the magnon quasi-particle operators $\beta_{\bd{k}}$ 
the Hamiltonian is diagonal,
 \begin{equation}
 {\cal{H}}_{\rm m}^{(2)} = \sum_{\bd{k}} 
 \left[ E_{\bd{k}} \beta^{\dagger}_{\bd{k}} \beta_{\bd{k}} + 
 \frac{ A_{\bd{k}} - E_{\bd{k}}}{2 E_{\bd{k}} }
 \right].
 \label{eq:Hmbog}
 \end{equation}
For a thin stripe with thickness $d$ the magnon dispersion
can be approximated for small wave vectors by~\cite{Kreisel09,Kalinikos86}
 \begin{eqnarray}
 E_{\bd{k}} & = & [ h + \rho_{s} \bd{k}^2 +
 \Delta ( 1 - f_{\bd{k}} ) \sin^2 \theta_{\bd{k}}  ]^{1/2}
 \nonumber
 \\
 &  \times &  [ h + \rho_{s} \bd{k}^2 +
  \Delta f_{\bd{k}}  ]^{1/2},
 \label{eq:Eklong}
 \end{eqnarray}
where $\theta_{\bd{k}}$ is the angle between the magnetic field
and the wave vector $\bd{k}$ and
the form factor can be approximated by
 \begin{equation}
 f_{\bd{k}} \approx \frac{ 1 - e^{ - | \bd{k} | d }}{ |  \bd{k} | d }.
 \end{equation}
For a thin YIG stripe the exchange spin stiffness  $\rho_{s}$ and the
dipolar energy scale $\Delta$ have the values \cite{Gurevich96,Gilleo58}
 \begin{eqnarray}
 \rho_{s}  & = & 2 \mu_B \times 5.17 \times 10^{-9} \, {\rm Oe} ~{\rm cm^2}   
 \nonumber
 \\
 & \approx  & 945  \,  {\rm GHz} \times a^2 \times 2 \pi \hbar,
 \\
 \Delta & = & 3 S D^{zz}_0 = 2 \mu_B \times 4 \pi  M_s  
  = 2 \mu_B \times 1750 \, {\rm G}
 \nonumber
 \\
 & \approx &  0.235 \, {\rm K} \times k_B =  4.89 \,  {\rm GHz} \times 2 \pi \hbar.
 \end{eqnarray}   
Note that in units where $k_B = 1$ and $h = 2 \pi \hbar =1$
we have $1 \, {\rm K} \approx 20.8  \, {\rm GHz}$.

\section{Magnon-phonon Hamiltonian of thin YIG-films}
\label{sec:magpho}

One source of the spin-phonon coupling is the dependence
of the true positions $\bd{r}_i = \bd{R}_i + \bd{X}_i$
of the spins on the phonon displacements
$\bd{X}_i = \bd{X} ( \bd{R}_i )$ at the lattice sites
$\bd{R}_i$. The resulting magnon phonon interaction
can be derived from the effective spin-model (\ref{eq:Hspin})
by expanding the exchange couplings $J_{ij} = J ( \bd{R}_i - \bd{R}_i
 + {\bd{X}}_i - \bd{X}_j )$ in powers of the phonon 
displacements \cite{Kreisel11}.
However, in collinear magnets such a procedure
does not take into account the dominant source of the
magnon-phonon interaction, which is generated by
relativistic effects such as dipole-dipole interactions 
and spin-orbit coupling \cite{Gurevich96}. These effects
involve the charge degrees of freedom so that they
cannot be simply included in our effective spin model (\ref{eq:Hspin}).
To derive the proper quantized interaction
between magnons and phonons in YIG,
we therefore follow the semi-phenomenological approach
pioneered by Abrahams and Kittel \cite{Abrahams52}, which relies
on the quantization of the phenomenological expression for the
classical magneto-elastic energy.

\subsection{Classical magneto-elastic energy in YIG}

To  second order in the
spin-density field $\bd{M} ( \bd{r} )$ and to
first order in the displacement field 
 $\bd{X} ( \bd{r} )$ associated with long-wavelength
acoustic phonons 
the phenomenological expression for the
magneto-elastic energy is \cite{Abrahams52,Kittel58,Kaganov59,Lord68,Gurevich96}
\begin{eqnarray}
E_{\rm me} [ \bd{M},    \bd{X} ] & = & \frac{n}{M^2_s}
 \int d^3 r  \sum_{\alpha \beta }
 \Biggl[
 B_{\alpha \beta } M_{\alpha} ( \bd{r} ) M_{\beta} ( \bd{r} )
 \nonumber
 \\
 & &    + 
 B^{\prime}_{\alpha \beta }
 \frac{ \partial \bd{M} ( \bd{r} ) }{ \partial r_{\alpha} }
 \cdot
\frac{ \partial \bd{M} ( \bd{r} ) }{ \partial r_{\beta} }
  \Biggr]X_{\alpha \beta } ( \bd{r} )    ,
  \label{eq:Eme}
  \hspace{7mm}
 \end{eqnarray}
where 
 $\bd{M} ( \bd{r} )$ is the local magnetization
(i.e., the magnetic moment per unit volume at position $\bd{r}$),
$M_s$ is the saturation magnetization,  
$n = N / V$ is the number density of the magnetic particles
in the system, and 
 \begin{equation}
 X_{\alpha \beta } ( \bd{r} ) = \frac{1}{2} \left[
 \frac{ \partial X_{\alpha} ( \bd{r} )}{\partial r_{\beta} }
 + \frac{\partial X_{\beta} ( \bd{r} )}{\partial r_{\alpha} }
 \right]
 \end{equation}
is the symmetric strain tensor.
Here $X_{\alpha} ( \bd{r} ) = \bd{e}_{\alpha} \cdot \bd{X} ( \bd{r} )$
are the projections
of the phonon displacement field $\bd{X} ( \bd{r} )$
onto the cartesian unit vectors $\bd{e}_{\alpha}$, where $\alpha =x,y,z$.
The  first term on the right-hand side of
Eq.~(\ref{eq:Eme}) involving the
couplings  $B_{\alpha \beta }$
is due to
relativistic effects such as dipole-dipole interactions 
and spin-orbit coupling, while the gradient term
involving $B^{\prime}_{\alpha \beta}$
is generated by the dependence of the exchange interaction on the 
phonon coordinates.
For a cubic lattice the phenomenological
coupling tensors $B_{\alpha \beta}$ and $B^{\prime}_{\alpha \beta}$ have the
structure
 \begin{subequations}
\begin{eqnarray}
 B_{\alpha \beta } & = &  \delta_{\alpha \beta } B_{\parallel} + (1 - \delta_{\alpha \beta } ) B_{\bot},
 \\
 B^{\prime}_{\alpha \beta } & = &  
 \delta_{\alpha \beta } B^{\prime}_{\parallel} + 
 (1 - \delta_{\alpha \beta } ) B^{\prime}_{\bot}.
 \end{eqnarray}
 \end{subequations}
The prefactor of $n / M_s^2$ 
in Eq.~(\ref{eq:Eme}) is introduced such
that $B_{\bot}$ and $B_{\parallel}$ have units of energy.
Transforming to wave vector space,
 \begin{subequations}
 \begin{eqnarray}
 \bd{M} ( \bd{r} ) & = & \int_{\bd{k}} e^{ i \bd{k} \cdot
 {\bd{r} } } \bd{M} ( \bd{k} ),
 \\
  \bd{X} ( \bd{r} ) & = &  \int_{\bd{k}}   e^{ i \bd{k} \cdot
 {\bd{r} } } \bd{X} ( \bd{k} ),
 \\
  {X}_{\alpha \beta} ( \bd{r} ) & = &  \int_{\bd{k}}   e^{ i \bd{k} \cdot
 {\bd{r} } } {X}_{\alpha \beta} ( \bd{k} ),
 \end{eqnarray}
 \end{subequations}
where $\int_{\bd{k}} = \int \frac{d^3 k}{ ( 2 \pi )^3 }$,
the matrix elements of the strain tensor in Fourier space are
 \begin{equation}
 {X}_{\alpha \beta} ( \bd{k} ) = \frac{i}{2} \left[
 X_{\alpha} ( \bd{k} ) k_{\beta} + X_{\beta} ( \bd{k} )  k_{\alpha}
 \right],
 \end{equation}
and the magneto-elastic energy can be written as
 \begin{eqnarray}
 E_{\rm me} [ \bd{M},   \bd{X} ] & = & \frac{n}{M_s^2} 
\int_{\bd{k}} \int_{\bd{k}^{\prime}} 
 \int_{\bd{q}}
 ( 2 \pi )^3 \delta ( \bd{k} + \bd{k}^{\prime}+ \bd{q}  )
 \nonumber
 \\
 & \times &
 \sum_{\alpha \beta } \Bigl[
 B_{\alpha \beta } M_{\alpha} ( \bd{k} ) M_{\beta} ( \bd{k}^{\prime} )
 \nonumber
 \\
 & & \hspace{-0mm} 
  - B^{\prime}_{\alpha \beta }  k_{\alpha} k^{\prime}_{\beta } 
  \bd{M} ( \bd{k} )  \cdot \bd{M} ( \bd{k}^{\prime} )
  \Bigr]   X_{\alpha \beta } ( \bd{q} )   .
 \hspace{7mm}
 \end{eqnarray}

At long wavelength the dominant contribution to the 
 magneto-elastic coupling is due to spin-orbit coupling \cite{Gurevich96} so that
in this work we shall neglect the exchange contribution, setting
$B^{\prime}_{\alpha \beta } =0$.
For YIG at room temperature the numerical values for the 
magneto-elastic coupling constants are \cite{Gurevich96,Eggers63,Hansen73}
 \begin{equation}
 n B_{\parallel} = 3.48 \times 10^6 \, {\rm erg} /  {\rm cm}^3,
 \; \; \;
 n B_{\bot} = 6.96 \times 10^6 \, {\rm erg} / {\rm cm}^3.
 \end{equation}
The number density of the magnetic ions in YIG is \cite{Gilleo58}
  \begin{equation}
 n = 1/ a^3 ,
 \; \; \; a = 12.376\, {\rm \AA},
 \end{equation}
so that in units where the Boltzmann constant and $2 \pi \hbar$ are 
set equal to unity
\begin{equation}
 B_{\parallel} = 47.8 \, {\rm K} = 994 \, {\rm GHz},
 \; \; \;
 B_{\bot} = 95.6 \, {\rm K} = 1988 \, {\rm GHz}. 
 \end{equation}

\subsection{Quantized magnon-phonon interaction in the
Holstein-Primakoff basis}

Let us now quantize the magneto-elastic energy.
For the phonon field we adopt the usual
strategy of expressing the displacement field
by the bosonic creation and 
annihilation operators $a^{\dagger}_{\bd{k} \lambda}$ and $a_{\bd{k} \lambda}$ of the phonon eigenmodes
with momentum $\bd{k}$ and polarization $ \bd{e}_{\bd{k} \lambda}$,
 \begin{eqnarray}
 \bd{X} ( \bd{k} )   & \rightarrow  & \frac{V}{\sqrt{N} } \bd{X}_{\bd{k}},
 \\
 \bd{X}_{\bd{k}} & = & \sum_{\lambda} X_{\bd{k} \lambda}
  \bd{e}_{\bd{k} \lambda}
 =
 \sum_{\lambda}
  \frac{ a_{\bd{k} \lambda} + 
 a^{\dagger}_{-\bd{k} \lambda}     }{ \sqrt{ 2 m
 \omega_{\bd{k} \lambda} }}   \bd{e}_{\bd{k} \lambda}  ,
 \label{eq:Xdef}
 \end{eqnarray}
where $\lambda = 1,2,3$ labels the three acoustic phonon branches, 
$\omega_{\bd{k} \lambda} = c_{\lambda} | \bd{k} |$
are the phonon dispersions, $m$ is the
effective ionic mass in a unit cell,
$V$ is the volume of the system, and $N$ is the number of unit cells. 
Note that in general
the polarization vectors satisfy~\cite{footnotepol} the orthogonality
relations 
$\bd{e}_{\bd{k} \lambda}^{\ast} \cdot \bd{e}_{\bd{k} \lambda^{\prime} } =
 \delta_{\lambda \lambda^{\prime}}$ and the completeness relation
$\sum_{\lambda} \bd{e}_{\bd{k} \lambda} \bd{e}_{\bd{k} \lambda}^{\dagger}
 = \mathbf{1}$,
so that we may identify
 \begin{equation}
 X_{\bd{k} \lambda} =
 \bd{e}_{ \bd{k} \lambda}^{\ast} \cdot \bd{X}_{\bd{k}} = 
  \frac{ a_{\bd{k} \lambda} + 
 a^{\dagger}_{-\bd{k} \lambda}     }{ \sqrt{ 2 m
 \omega_{\bd{k} \lambda} }}.
 \label{eq:Xkdef}
 \end{equation}
Below we shall denote the Fourier transform of the quantized strain tensor by
 \begin{equation}
 X_{\bd{k}}^{\alpha \beta } = \frac{i}{2} \bd{k}_{\alpha \beta}
 \cdot  \bd{X}_{\bd{k}} ,
 \; \; \; \bd{k}_{\alpha \beta} =  k_{\alpha}   \bd{e}_{\beta} +
  k_{\beta} \bd{e}_{\alpha}.
 \end{equation}
We may also choose the phases of the unit vectors such that \cite{footnotepol}
$\bd{e}_{\bd{k} \lambda}^{\ast} = \bd{e}_{- \bd{k} \lambda }$.
For a quasi two-dimensional YIG stripe in the $yz$-plane
we parametrize the in-plane wave vectors as
% \begin{equation}
 $\bd{k} = | \bd{k} | ( \cos \theta_{\bd{k}}  \bd{e}_z 
 + \sin \theta_{\bd{k}}  \bd{e}_y )$,
% \end{equation}
as shown in Fig.~\ref{fig:geometry}.
A convenient choice for the 
polarization vectors of the longitudinal and transverse
phonons is then
 \begin{subequations}
 \begin{eqnarray}
 \bd{e}_{ \bd{k} \parallel} & = &  i (   \bd{e}_z   \cos \theta_{\bd{k}} 
 +   \bd{e}_y \sin \theta_{\bd{k}}  ), 
 \label{eq:epar}
 \\
 \bd{e}_{ \bd{k} \bot 1} & = &  i ( \bd{e}_z   \sin \theta_{\bd{k}}  
  -  \bd{e}_y   \cos \theta_{\bd{k}}  ), 
 \label{eq:ebot1}
 \\
  \bd{e}_{ \bd{k} \bot 2} & = & \bd{e}_x.
 \label{eq:ebot2}
 \end{eqnarray}
 \end{subequations}
For YIG the relevant value of the effective ionic mass
appearing in Eqs.~(\ref{eq:Xdef}, \ref{eq:Xkdef}) is~\cite{Gurevich96,Gilleo58}
 \begin{equation}
 m = \rho a^3 , \; \; \; \rho  = 5.17 \, {\rm g}/{\rm cm}^3,
 \end{equation}
and the longitudinal and transverse phonon velocities are
 \begin{equation}
 c_{\parallel} = 7.209 \times 10^5 \, {\rm cm}/ {\rm s},
 \; \; \; 
 c_{\bot} = 3.843 \times 10^5 \, {\rm cm}/ {\rm s}.
 \end{equation}

To quantize the magnetic sector,
we follow Refs.[\onlinecite{Abrahams52,Kittel58,Kaganov59}]  
and replace the magnetization $ \bd{M} ( \bd{r} )$ at lattice site
$\bd{r} = \bd{R}_i$
by the corresponding spin-operator $\bd{S}_i$ according to the prescription
 \begin{equation}
 \bd{M} ( \bd{R}_i )  \rightarrow  2 \mu_B  n \bd{S}_i ,
 \label{eq:mri}
 \end{equation}
where $n = N / V = 1/ a^3$ is the number density.
In Fourier space Eq.~(\ref{eq:mri}) corresponds to
 \begin{equation}
 \bd{M} ( \bd{k} ) \rightarrow g \mu_B \sqrt{N} \bd{S}_{\bd{k}} =
 g \mu_B 
 \sum_{i} e^{ - i \bd{k} \cdot \bd{R}_i } \bd{S}_i ,
 \label{eq:FTspin}
 \end{equation}
Assuming that the macroscopic magnetization points
along the $z$ axis, we may express the
components of the spin operators at lattice site
$\bd{R}_i$ in terms of Holstein-Primakoff bosons (magnons) $b_i$ and $b_i^{\dagger}$ 
as usual with Eqs.~(\ref{eq:hp1})--(\ref{eq:hp3}).
In momentum space we have 
 \begin{equation}
 S^z_{ {\bd{k}} }  = \sqrt{N} S \delta_{\bd{k} , 0 } - \frac{1}{\sqrt{N}}
 \sum_{\bd{q}} b^{\dagger}_{\bd{q}} b_{\bd{q} + \bd{k}},
 \end{equation}
and to leading order in an expansion in powers of $1/S$,
 \begin{eqnarray}
 S^x_{\bd{k}}  & \approx &  \frac{  \sqrt{2  S}  }{2} ( b_{\bd{k}} +
 b^{\dagger}_{- \bd{k}} ),
 \\
 S^y_{\bd{k}} & \approx & \frac{ \sqrt{2S}  }{2i} ( b_{\bd{k}} -
 b^{\dagger}_{- \bd{k}} ).
 \end{eqnarray}
For large $S$ it is reasonable to retain only terms up to 
quadratic order in the 
magnons, so that the resulting
magneto-elastic Hamiltonian can be written as
 \begin{equation}
 {\cal{H}}_{\rm me} = {\cal{H}}_{{\rm me}}^{(2)}+ {\cal{H}}_{{\rm me}}^{(3)},
 \end{equation}
where the superscript indicates the number of operators involved.
The quadratic  term  ${\cal{H}}_{{\rm me}}^{(2)}$ 
is
 \begin{eqnarray}
 {\cal{H}}_{\rm me}^{(2)} & = & B_{\bot} \sqrt{ \frac{2}{S}} 
 \sum_{\bd{k}} 
 \Bigl[ X_{-\bd{k}}^{ zx}  ( b_{\bd{k}} + b^{\dagger}_{ -\bd{k}} ) 
% \nonumber
% \\
% & & \hspace{17mm}
 - i  X_{- \bd{k}}^{ zy}  ( b_{\bd{k}} - b^{\dagger}_{ - \bd{k}} ) 
 \Bigr]
 \nonumber
 \\
 & = & 
\sum_{\bd{k}}  \bd{X}_{-\bd{k}} \cdot
 \left[
 \bd{\Gamma}_{\bd{k}} 
 b_{\bd{k}}
 + \bd{\Gamma}^{\ast}_{- \bd{k}}  b^{\dagger}_{-\bd{k}} 
 \right]
 \nonumber
 \\
 & = & 
\sum_{\bd{k}} 
 \left[ \bd{X}_{\bd{k}}^{\ast} \cdot
 \bd{\Gamma}_{\bd{k}} b_{\bd{k}} +
 \bd{X}_{\bd{k}} \cdot
 \bd{\Gamma}_{\bd{k}}^{\ast} b^{\dagger}_{\bd{k}} 
 \right] ,
 \label{eq:H2}
 \end{eqnarray}
where we have used the fact that the $\bd{X}_{\bd{k}}$ 
are the Fourier components of a hermitian operator
so that
 \begin{equation}
 \bd{X}_{- \bd{k} } = \bd{X}_{\bd{k}}^{\ast},
 \end{equation}
and the vector vertex $\bd{\Gamma}_{\bd{k}}$ is defined by
 \begin{eqnarray}
 \bd{\Gamma}_{\bd{k}}  & =  &   
 - \frac{  B_{\bot}}{ \sqrt{2S}} 
 \left( \bd{k}_{yz} + i \bd{k}_{xz} \right)
 \nonumber
 \\
 & = &
- \frac{  B_{\bot}}{ \sqrt{2S}} 
   \left[ ( k_y + i k_x ) \bd{e}_z  +  k_z ( \bd{e}_y + i \bd{e}_x )   \right],
 \end{eqnarray}
Expanding
 \begin{equation}
  \bd{X}_{\bd{k}}  =  \sum_{\lambda} X_{\bd{k} \lambda}
  \bd{e}_{\bd{k} \lambda}, \; \; \;
\bd{\Gamma}_{\bd{k}}  =  \sum_{\lambda} \Gamma_{\bd{k} \lambda}
  \bd{e}_{\bd{k} \lambda},
 \end{equation}
with
 \begin{equation}
  X_{\bd{k} \lambda} = \bd{e}^{\ast}_{\bd{k} \lambda}
 \cdot  \bd{X}_{\bd{k}}, \; \; \; 
 \Gamma_{\bd{k} \lambda} = \bd{e}^{\ast}_{\bd{k} \lambda}
 \cdot  \bd{\Gamma}_{\bd{k}},
 \end{equation}
the Hamiltonian (\ref{eq:H2}) can be written as
 \begin{eqnarray}
  {\cal{H}}_{\rm me}^{(2)} & = & \sum_{\bd{k} \lambda}
 \left[ \Gamma_{\bd{k} \lambda} X^{\ast}_{\bd{k} \lambda} b_{\bd{k}}
 +  \Gamma_{\bd{k} \lambda}^{\ast} X_{\bd{k} \lambda} b^{\dagger}_{\bd{k}}
 \right]
 \nonumber
 \\
 & = & \sum_{\bd{k} \lambda} X_{- \bd{k} \lambda} 
 \left[ \Gamma_{\bd{k} \lambda} b_{\bd{k}}
 +  \Gamma_{-\bd{k} \lambda}^{\ast} b^{\dagger}_{-\bd{k}}
 \right].
 \label{eq:H2b}
 \end{eqnarray}
This part of the magneto-elastic Hamiltonian
describes the hybridisation between the phonon and magnon degrees of freedom.
With the choice (\ref{eq:epar})--(\ref{eq:ebot2}) of the
phonon basis we obtain for the projections of the hybridisation vertex
 \begin{subequations}
 \begin{eqnarray}
 \Gamma_{ \bd{k} \parallel} & = & i \frac{B_{\bot}}{ \sqrt{2S}}
  \frac{ 2 k_y k_z}{ | \bd{k} | } =  i \frac{B_{\bot}}{ \sqrt{2S}}
 | \bd{k} | \sin ( 2 \theta_{\bd{k}} ), 
 \label{eq:hybparallel}
 \\
 \Gamma_{ \bd{k} \bot 1} & = & i \frac{B_{\bot}}{ \sqrt{2S}}
  \frac{ k^2_y - k^2_z}{ | \bd{k} | } =  - i \frac{B_{\bot}}{ \sqrt{2S}}
 | \bd{k} | \cos ( 2 \theta_{\bd{k}} ), 
 \hspace{7mm}
 \\
 \Gamma_{ \bd{k} \bot 2} & = & - i \frac{B_{\bot}}{ \sqrt{2S}}
  k_z =  - i \frac{B_{\bot}}{ \sqrt{2S}}
 | \bd{k} | \cos   \theta_{\bd{k}} . 
 \end{eqnarray}
 \end{subequations}

Before discussing the spectrum of the magneto-elastic modes,
let us write down 
the cubic term  ${\cal{H}}_{{\rm me}}^{(3)}$  of the magneto-elastic
Hamiltonian in the form \cite{Kreisel11}
\begin{eqnarray}
& &  {\cal{H}}_{\rm me}^{(3)}  =  \frac{1}{\sqrt{N}} \sum_{\bd{k}  \bd{k}^{\prime}}
 \Bigl[ \bd{ \Gamma}^{ \bar{b} b}_{\bd{k}, \bd{k}^{\prime}} 
 \cdot \bd{X}_{\bd{k} - \bd{k}^{\prime}} b^{\dagger}_{\bd{k}} b_{\bd{k}^{\prime}}
 \nonumber
 \\
 & &  + \frac{1}{2!} 
 \left(  
 \bd{ \Gamma}^{ b b}_{  \bd{k} , \bd{k}^{\prime}} 
 \cdot \bd{X}_{- \bd{k} - \bd{k}^{\prime}} b_{\bd{k}} 
 b_{\bd{k}^{\prime}} 
 +
\bd{ \Gamma}^{ \bar{b} \bar{b}}_{\bd{k}, \bd{k}^{\prime}} 
 \cdot \bd{X}_{\bd{k} + \bd{k}^{\prime}} b^{\dagger}_{\bd{k}} 
 b^{\dagger}_{\bd{k}^{\prime}}  
\right) \Bigr]
\nonumber
 \\
 &= &  \frac{1}{\sqrt{N}} \sum_{\bd{k}  \bd{q}}
 \bd{X}_{-\bd{q}} \cdot
 \Bigl[ \bd{ U}_{- \bd{q}} 
  b^{\dagger}_{\bd{k}} b_{\bd{k} + \bd{q}}
 \nonumber
 \\
 & & \hspace{14mm}
 + \frac{1}{2!} \left(
\bd{ V}_{  - \bd{q}} 
 b_{-\bd{k}} 
 b_{ \bd{k}   + \bd{q} } 
+
   \bd{ V}^{\ast}_{ \bd{q}} 
 b^{\dagger}_{\bd{k}} 
 b^{\dagger}_{-\bd{k} - \bd{q}}
 \right) \Bigr]
 ,
 \label{eq:H3}
 \end{eqnarray} 
where the vector vertices are given by 
 \begin{subequations}
 \begin{eqnarray}
   \bd{ \Gamma}^{ \bar{b} b}_{\bd{k}, \bd{k}^{\prime}} & = & \bd{U}_{  \bd{k} - \bd{k}^{\prime}},
 \\
  \bd{ \Gamma}^{ b b}_{\bd{k}, \bd{k}^{\prime}} & = &\bd{V}_{  \bd{k} +  \bd{k}^{\prime}},
 \\
 \bd{ \Gamma}^{ \bar{b} \bar{b}}_{\bd{k}, \bd{k}^{\prime}} & = &
 \bd{V}^{\ast}_{  - \bd{k} -  \bd{k}^{\prime}} ,
 \end{eqnarray}
 \end{subequations}
with
 \begin{subequations}
 \begin{eqnarray}
  \bd{U}_{\bd{q}}
   &  = & 
 \frac{i B_{\parallel}}{ S}
 ( q_x \bd{e}_x + q_y \bd{e}_y - 2 q_z \bd{e}_z ),
 \\
  \bd{V}_{\bd{q}}  & = &
 \frac{i B_{\parallel}}{ S} ( q_x \bd{e}_x - q_y \bd{e}_y )
 + \frac{ B_{\bot}}{ S} \bd{q}_{xy}
 \nonumber
 \\
 & =  &
 \frac{i B_{\parallel}}{ S} ( q_x \bd{e}_x - q_y \bd{e}_y )
 + \frac{ B_{\bot}}{ S} ( q_x \bd{e}_y + q_y \bd{e}_x ).
 \hspace{7mm}
 \end{eqnarray}

 \end{subequations}

\subsection{Magneto-elastic interaction vertices in the
quasi-particle basis}

For the calculation of the spectral function of the magnons 
it is useful to express the magneto-elastic interaction
in terms of the magnon quasi-particle operators $\beta_{\bd{k}}$ and
$\beta_{\bd{k}}^{\dagger}$ which are related to the
Holstein-Primakoff bosons $b_{\bd{k}}$ and $b_{\bd{k}}^{\dagger}$
via the Bogoliubov transformation (\ref{eq:Bogo}).
Substituting this transformation into Eqs.~(\ref{eq:H2}) and (\ref{eq:H3})
we obtain for the hybridisation part
 \begin{eqnarray}
  {\cal{H}}_{\rm me}^{(2)} & = & 
\sum_{\bd{k} \lambda} X_{- \bd{k} \lambda} 
 \left[ {\Gamma}^{\beta}_{\bd{k} \lambda} \beta_{\bd{k}}
 +  {\Gamma}^{\bar{\beta}}_{-\bd{k} \lambda} \beta^{\dagger}_{-\bd{k}}
 \right],
 \label{eq:H2beta}
 \end{eqnarray}
with
 \begin{eqnarray}
 {\Gamma}^{\beta}_{\bd{k} \lambda} & = & 
 u_{\bd{k}} \Gamma_{\bd{k} \lambda} - v_{\bd{k}}^{\ast} 
 \Gamma^{\ast}_{ - \bd{k} \lambda },
 \\
 {\Gamma}^{\bar{\beta}}_{\bd{k} \lambda} & = & 
 u_{\bd{k}} \Gamma^{\ast}_{\bd{k} \lambda} - v_{\bd{k}} 
 \Gamma_{ - \bd{k} \lambda } =  ( {\Gamma}^{\beta}_{\bd{k} \lambda})^{\ast}  .
 \end{eqnarray}
The magnon-phonon interaction defined in Eq.~(\ref{eq:H3})
can be written as
\begin{eqnarray}
& &  {\cal{H}}_{\rm me}^{(3)}  =  \frac{1}{\sqrt{N}} \sum_{\bd{k}  \bd{k}^{\prime}}
 \Bigl[ \bd{ \Gamma}^{ \bar{\beta} \beta}_{\bd{k}, \bd{k}^{\prime}} 
 \cdot \bd{X}_{\bd{k} - \bd{k}^{\prime}} \beta^{\dagger}_{\bd{k}} \beta_{\bd{k}^{\prime}}
 \nonumber
 \\
 & &  + \frac{1}{2!} 
 \left(  
 \bd{ \Gamma}^{ \beta \beta}_{  \bd{k} , \bd{k}^{\prime}} 
 \cdot \bd{X}_{- \bd{k} - \bd{k}^{\prime}} \beta_{\bd{k}} 
 \beta_{\bd{k}^{\prime}} +
\bd{ \Gamma}^{ \bar{\beta} \bar{\beta}}_{\bd{k}, \bd{k}^{\prime}} 
 \cdot \bd{X}_{\bd{k} + \bd{k}^{\prime}} \beta^{\dagger}_{\bd{k}} 
 \beta^{\dagger}_{\bd{k}^{\prime}}  
\right) \Bigr],
 \nonumber
 \\
 & &
\end{eqnarray}
with
\begin{subequations}
 \begin{align}
 {\bd{\Gamma}}^{ \bar{\beta} \beta}_{ \bd{k} , \bd{k}^{\prime}} & = 
u_{\bd{k}} u_{\bd{k}^{\prime}}
 {\bd{\Gamma}}^{ \bar{b}b }_{ \bd{k} , \bd{k}^{\prime}}
+ v_{\bd{k}} v^{\ast}_{\bd{k}^{\prime}}
 {\bd{\Gamma}}^{\bar{b} b }_{ - \bd{k}^{\prime} , - \bd{k}}
 \nonumber
 \\
 &  - v_{\bd{k}} u_{\bd{k}^{\prime}}
 {\bd{\Gamma}}^{bb }_{ - \bd{k} ,  \bd{k}^{\prime}} 
 -
   u_{\bd{k}} v^{\ast}_{\bd{k}^{\prime}}
 {\bd{\Gamma}}^{ \bar{b} \bar{b} }_{ \bd{k} , - \bd{k}^{\prime}}
 \nonumber
 \\
 & \hspace{-10mm} =  
 ( u_{\bd{k}} u_{\bd{k}^{\prime}}
+ v_{\bd{k}} v^{\ast}_{\bd{k}^{\prime}} ) \bd{U}_{\bd{k} - \bd{k}^{\prime} }
%\\
% & 
 - v_{\bd{k}} u_{\bd{k}^{\prime}} \bd{V}_{ \bd{k}^{\prime}  - \bd{k}  }
 - u_{\bd{k}} v^{\ast}_{\bd{k}^{\prime}} 
 \bd{V}^{\ast}_{\bd{k}^{\prime} - \bd{k}} ,
 \label{eq:Gammabb1}
 \\
 {\bd{\Gamma}}^{\beta \beta   }_{ \bd{k} , \bd{k}^{\prime}} & = 
 u_{\bd{k}} u_{\bd{k}^{\prime}} 
 {\bd{\Gamma}}^{b b   }_{ \bd{k} , \bd{k}^{\prime}}
+ v^{\ast}_{\bd{k}} v^{\ast}_{\bd{k}^{\prime}}
 {\bd{\Gamma}}^{ \bar{b} \bar{b}   }_{ - \bd{k} , - \bd{k}^{\prime}}
 \nonumber
 \\
 & -   v^{\ast}_{\bd{k}} u_{\bd{k}^{\prime}} {\bd{\Gamma}}^{\bar{b}b }_{  
 - \bd{k},  \bd{k}^{\prime} }
-  u_{\bd{k}} v^{\ast}_{\bd{k}^{\prime}} 
 {\bd{\Gamma}}^{\bar{b}b }_{  -\bd{k}^{\prime} ,  \bd{k}}
 \nonumber
 \\
 &  \hspace{-10mm} =  u_{\bd{k}} u_{\bd{k}^{\prime}} 
 {\bd{V}}_{ \bd{k} + \bd{k}^{\prime}}
 + v^{\ast}_{\bd{k}} v^{\ast}_{\bd{k}^{\prime}}
 {\bd{V}}^{\ast}_{  \bd{k} +  \bd{k}^{\prime}}
 - (  v^{\ast}_{\bd{k}} u_{\bd{k}^{\prime}} + 
 u_{\bd{k}} v^{\ast}_{\bd{k}^{\prime}} ) \bd{U}_{ - \bd{k} - \bd{k}^{\prime}},
 \label{eq:Gammabb2}
 \\
{\bd{\Gamma}}^{ \bar{\beta} \bar{\beta}   }_{ \bd{k} , \bd{k}^{\prime}}
& = 
 u_{\bd{k}} u_{\bd{k}^{\prime}} 
{\bd{\Gamma}}^{\bar{b} \bar{b}   }_{ \bd{k} , \bd{k}^{\prime}}
+ v_{\bd{k}} v_{\bd{k}^{\prime}}
 {\bd{\Gamma}}^{b b   }_{ - \bd{k} , - \bd{k}^{\prime}}
 \nonumber
 \\
 & - 
  u_{\bd{k}} v_{\bd{k}^{\prime}} {\bd{\Gamma}}^{\bar{b}b }_{  \bd{k} ,  -\bd{k}^{\prime}}
  -  v_{\bd{k}} u_{\bd{k}^{\prime}} {\bd{\Gamma}}^{\bar{b}b }_{   \bd{k}^{\prime}, -\bd{k} }
 \nonumber
 \\
 & \hspace{-10mm} =   u_{\bd{k}} u_{\bd{k}^{\prime}} 
 {\bd{V}}^{\ast}_{ - \bd{k} -  \bd{k}^{\prime}}
 + v_{\bd{k}} v_{\bd{k}^{\prime}}
 {\bd{V}}_{  - \bd{k} -  \bd{k}^{\prime}}
 - (  u_{\bd{k}} v_{\bd{k}^{\prime}} + 
 v_{\bd{k}} u_{\bd{k}^{\prime}} ) \bd{U}_{  \bd{k} + \bd{k}^{\prime}}.
\nonumber
 \\
 &
 \label{eq:Gammabb3}
 \end{align}
\end{subequations}

Note that the hermiticity of the Hamiltonian implies
 \begin{eqnarray}
 \bd{\Gamma}^{ \bar{\beta} \beta }_{ \bd{k} , \bd{k}^{\prime}}
 & = &  ( \bd{\Gamma}^{ \bar{\beta} \beta }_{ \bd{k}^{\prime} , \bd{k} } )^{\ast}
 \;  ,
 \; \; \; 
 \bd{\Gamma}^{ {\beta} \beta }_{ \bd{k} , \bd{k}^{\prime}}
  =   ( \bd{\Gamma}^{ \bar{\beta} \bar{\beta} }_{ \bd{k} , \bd{k}^{\prime}} )^{\ast}.
 \end{eqnarray}
Below we shall need these interaction vertices to calculate the
damping of the magnons in YIG due to the coupling to the phonons.
In fact, we shall need the  projections of the
three-legged vertices onto the phonon basis,
which we define by
 \begin{subequations}
 \begin{eqnarray}
 \Gamma^{\bar{\beta} \beta}_{ \bd{k} , \bd{k}^{\prime} , \lambda}
 & = &   \bd{e}^{\ast}_{ \bd{k}^{\prime} - \bd{k} , \lambda} \cdot
 \bd{\Gamma}^{\bar{\beta} \beta}_{ \bd{k} , \bd{k}^{\prime} }
 \\
\Gamma^{{\beta} \beta}_{ \bd{k} , \bd{k}^{\prime} , \lambda}
 & = &   \bd{e}^{\ast}_{ \bd{k}^{\prime} + \bd{k} , \lambda} \cdot
 \bd{\Gamma}^{{\beta} \beta}_{ \bd{k} , \bd{k}^{\prime} },
 \end{eqnarray}
 \end{subequations}
or in terms of shifted labels,
 \begin{subequations}
 \begin{eqnarray}
 \Gamma^{\bar{\beta} \beta}_{ \bd{k} , \bd{k} + \bd{q} , \lambda}
 & = &   \bd{e}^{\ast}_{ \bd{q} , \lambda} \cdot
 \bd{\Gamma}^{\bar{\beta} \beta}_{ \bd{k} , \bd{k} + \bd{q} }
 \\
\Gamma^{{\beta} \beta}_{ \bd{k} , - \bd{k} + \bd{q} , \lambda}
 & = &   \bd{e}^{\ast}_{  \bd{q} , \lambda} \cdot
 \bd{\Gamma}^{{\beta} \beta}_{ \bd{k} , - \bd{k} + \bd{q} }.
 \end{eqnarray}
 \end{subequations}
These vertices  should also be useful in microscopic
calculations of the non-equilibrium dynamics of magnons in YIG.
Note that in 
 Ref.~[\onlinecite{Hick12}] only the Cherenkov type of process described by
the vertex
 ${\bd{\Gamma}}^{\beta^{\dagger}\beta}_{ \bd{k} , \bd{k}^{\prime}}$
has been taken into account (however, using a simple phenomenological
expression for this vertex).
It should be interesting to repeat the 
calculations of Ref.~[\onlinecite{Hick12}] for the non-equilibrium 
magnon dynamics in YIG
using the more realistic magnon-phonon vertices given above.

\section{Magneto-elastic modes in YIG}
\label{sec:hybrid}

To calculate the energy dispersion of the
magneto-elastic modes, it is sufficient to
retain only terms which are quadratic in the magnon and phonon operators.
The Hamiltonian of the coupled magnon-phonon system can then be approximated 
by 
 \begin{equation}
 {\cal{H}}^{(2)} =  {\cal{H}}^{(2)}_{\rm m} +  {\cal{H}}^{(2)}_{\rm e}
 + {\cal{H}}^{(2)}_{\rm me},
 \end{equation}
where the quadratic spin wave part 
$ {\cal{H}}^{(2)}_{\rm m}$
is given in
Eqs.~(\ref{eq:Hm}) and (\ref{eq:Hmbog}), the
pure phonon part can be written as
 \begin{eqnarray}
  {\cal{H}}^{(2)}_{\rm e} & = & \sum_{\bd{k} \lambda}
 \left[ \frac{ P_{ - \bd{k} \lambda} P_{\bd{k} \lambda}}{2 m}
 + \frac{m}{2} \omega_{\bd{k} \lambda}^2 X_{ - \bd{k} \lambda}
 X_{\bd{k} \lambda} \right]
 \nonumber
 \\
 & = & \sum_{\bd{k} \lambda}
 \omega_{\bd{k} \lambda} \left[ a^{\dagger}_{\bd{k} \lambda}
 a_{\bd{k} \lambda} + \frac{1}{2} \right],
 \end{eqnarray}
and the magnon-phonon hybridization  ${\cal{H}}^{(2)}_{\rm me}$
is given in Eqs.~(\ref{eq:H2b}) and (\ref{eq:H2beta}).

\subsection{Effective magnon action and magnon self-energies}
To study the effect of the lattice vibrations on 
the spin excitations, it is convenient to use a functional
integral formulation of the coupled magnon-phonon system and 
integrate over the phonon degrees of freedom,
which in our approximation can be done exactly because
we have truncated the expansion (\ref{eq:Eme}) 
of the magneto-elastic energy
at the linear order in the phonon coordinates.
The magnon operators $\beta_{\bd{k}}$ and $\beta^{\dagger}_{\bd{k}}$
in the quasi-particle basis
should then be represented by complex fields
$\beta_K$ and $\bar{\beta}_K$ depending on momentum $\bd{k}$ and
bosonic Matsubara frequency $ i \omega $, which 
we collect into the label $K = ( \bd{k} , i \omega )$.
The resulting
Euclidean  effective  action of the magnons is of the form
%\vspace*{-.2cm}
 \begin{equation}
  S [ \bar{\beta} , \beta ] =  S_2 [ \bar{\beta} , \beta ] +
 S_3 [ \bar{\beta} , \beta ] + S_4 [ \bar{\beta} , \beta ],
 \end{equation}
where the Gaussian part is given by
%\vspace*{-.2cm}
 \begin{eqnarray}
 S_2 [ \bar{\beta} , \beta ] & = & 
- \frac{1}{T} \sum_K \Bigl\{ \left[ i \omega -
 E_{\bd{k}} - \Sigma_1 ( K ) \right] \bar{\beta}_K \beta_K
 \nonumber
 \\
 & & - \frac{1}{2}
 \left[ \Pi_1 ( K ) \beta_{-K } \beta_K + 
\Pi_1^{\ast} ( K ) \bar{\beta}_{K } \bar{\beta}_{-K}
 \right] \Bigr\}.
 \hspace{7mm}
 \end{eqnarray}
Here the normal and anomalous self-energies to first order in
the small parameter $1/S$ are given by
 \begin{eqnarray}
 \Sigma_1 ( K ) & = & - \sum_{\lambda} 
 \frac{ | {\Gamma}^{\beta}_{ \bd{k} \lambda} |^2}{m ( \omega^2 +
 \omega^2_{\bd{k} \lambda} ) } =  - \sum_{\lambda} 
  | {\Gamma}^{\beta}_{ \bd{k} \lambda} |^2 F_0 ({ K \lambda})
 , \label{eq:sigma1}
 \nonumber
 \\
 & &
 \\
  \Pi_1 ( K ) & = & - \sum_{\lambda} 
 \frac{ {\Gamma}^{\beta}_{ \bd{k} \lambda}  
{\Gamma}^{\beta}_{ - \bd{k} \lambda}   }{m ( \omega^2 +
 \omega^2_{\bd{k} \lambda} ) }  =
 - \sum_{\lambda} 
{\Gamma}^{\beta}_{ \bd{k} \lambda}  
 {\Gamma}^{\beta}_{ - \bd{k} \lambda}  F_0 ({ K \lambda})
 , \label{eq:pi1}
 \nonumber
 \\
 & &
 \end{eqnarray}
where we have introduced the symmetric
phonon propagator
 \begin{equation}
 F_0 ( { K \lambda} ) = \frac{1}{ m ( \omega^2 +
 \omega^2_{\bd{k} \lambda} ) }.
 \end{equation}
The interference of the
magnon-phonon hybridisation with the 
cubic term of the magneto-elastic coupling yields
a cubic contribution to the effective magnon-magnon interaction,
\begin{eqnarray}
 S_3 [ \bar{\beta} , \beta ] & = & 
 - \frac{1}{T \sqrt{N}} \sum_{ K_1 K_2 K_3 } 
 \delta_{ K_1 + K_2 + K_3 ,0}
 \Bigr[ 
 \nonumber
 \\
 & & \hspace{-9mm}
\frac{1}{2} \Gamma^{ \bar{\beta} \beta \beta}_{ 1; 2 3 }   
 \bar{\beta}_{ -1} \beta_{ 2 } \beta_{3 } 
 +  \frac{1}{2} \Gamma^{ \bar{\beta} \bar{\beta}\beta}_{ 1 2 ; 3 }   
 \bar{\beta}_{ -1} \bar{\beta}_{ -2 } \beta_{3 } 
 \nonumber
 \\
 & & \hspace{-11mm} +
  \frac{1}{3!} \Gamma^{ \beta \beta \beta}_{ 1 2 3 }   
 \beta_{ 1} \beta_{ 2 } \beta_{3 } +  \frac{1}{3!} 
 \Gamma^{\bar{\beta} \bar{\beta} \bar{\beta}}_{123 }   
 \bar{\beta}_{ -1} \bar{\beta}_{ -2 } \bar{\beta}_{- 3 } 
 \Bigr],
 \label{eq:S3def}
\end{eqnarray}
Here $\delta_{ K , K^{\prime} } = \delta_{\bd{k} , \bd{k}^{\prime}}
 \delta_{ \omega , \omega^{\prime}}$ and for
simplicity we have 
abbreviated $\beta_1 \equiv  \beta_{ K_1 } $ and 
$\Gamma^{\bar{\beta} \beta \beta}_{K_1 ;
 K_2 K_3} \equiv \Gamma^{\bar{\beta}\beta \beta}_{ 1 ;
 2 3}$, and similarly for the other labels.
Introducing the notation $F_0 ( K_1 \lambda ) = F_{1 \lambda}$,
the properly symmetrized cubic interaction vertices 
can be written as
 \begin{widetext}
 \begin{subequations}
 \begin{eqnarray}
 \Gamma^{ \bar{\beta} \beta \beta}_{ 1; 2 3 }   & = &
 \sum_{\lambda} \Bigl[
 F_{ 1 \lambda}   {\Gamma}^{\bar{\beta}}_{-1 \lambda}
 \Gamma^{\beta \beta}_{ 2 , 3 , \lambda}
 +  F_{ 2 \lambda}  {\Gamma}^{\beta}_{2 \lambda}
  \Gamma^{\bar{\beta} \beta}_{ - 1 , 3 , \lambda}
% \nonumber
% \\
%  &   & \hspace{5mm} 
 + F_{ 3 \lambda} {\Gamma}^{\beta}_{3 \lambda}
 \Gamma^{\bar{\beta} \beta}_{ - 1 , 2 , \lambda}
 \Bigr],
 \label{eq:Ga1}
 \\
 \Gamma^{ \bar{\beta} \bar{\beta}\beta}_{ 1 2 ; 3 }  & = &
 \sum_{\lambda}
 \Bigl[
  F_{ 1 \lambda}  {\Gamma}^{\bar{\beta}}_{-1 \lambda} 
  \Gamma^{\bar{\beta} \beta}_{ - 2 , 3 , \lambda}
 +
 F_{ 2 \lambda} {\Gamma}^{\bar{\beta}}_{-2 \lambda}
  \Gamma^{\bar{\beta} \beta}_{ - 1 , 3 , \lambda}
% \nonumber
% \\
% &+ &  
 + F_{ 3 \lambda} {\Gamma}^{\beta}_{3 \lambda}
 \Gamma^{\bar{\beta} \bar{\beta}}_{ -1 , -2 , \lambda} \Bigr],
 \\
 \Gamma^{ \beta \beta \beta}_{ 1 2 3 }  & = &
 \sum_{\lambda} \Bigr[
 F_{ 1 \lambda}{\Gamma}^{\beta}_{1 \lambda}
\Gamma^{{\beta} {\beta}}_{ 2 , 3 , \lambda}
 + 
(1 \leftrightarrow 2)+ (1  \leftrightarrow 3 ) \Bigr],
 \\
 \Gamma^{\bar{\beta} \bar{\beta} \bar{\beta}}_{ 1 2 3 }   & = &
 \sum_{\lambda} \Bigr[
 F_{ 1 \lambda} {\Gamma}^{\bar{\beta}}_{-1 \lambda}
\Gamma^{\bar{\beta} \bar{\beta}}_{ -2 , -3 , \lambda}
 +(1 \leftrightarrow 2)+ (1  \leftrightarrow 3 ) \Bigl].
 \hspace{7mm}
 \end{eqnarray}
 \end{subequations}
Finally, the quartic magnon-magnon interaction
is generated from the square of the cubic magneto-elastic coupling
via the exchange of a virtual phonon,
\begin{eqnarray}
 & & S_4 [ \bar{\beta} , \beta ]  =  
 - \frac{1}{T N} \sum_{ K_1 \ldots K_4} 
 \delta_{ K_1 + \ldots + K_4 ,0}
 \Bigr[ 
 \nonumber
 \\
 & &
\frac{1}{(2!)^2} \Gamma^{ \bar{\beta} \bar{\beta} \beta \beta}_{ 12; 3 4}   
 \bar{\beta}_{ -1} \bar{\beta}_{-2} \beta_{ 3 } \beta_{4 } 
 \nonumber
 \\
 & & +
\frac{1}{3!} \Gamma^{ \bar{\beta} \beta \beta \beta}_{ 1; 2 3 4}   
\bar{\beta}_{ -1} \beta_{ 2 } \beta_{3 } \beta_4
 +  \frac{1}{3!} \Gamma^{ \bar{\beta} \bar{\beta} \bar{\beta} 
 \beta}_{ 1 2 3; 4 }   
 \bar{\beta}_{ -1} \bar{\beta}_{ -2 } \bar{\beta}_{-3} \beta_{4 } 
 \nonumber
 \\
 & & +
  \frac{1}{4!} \Gamma^{\beta \beta \beta \beta}_{ 1 2 3 4}   
\beta_{ 1} \beta_{ 2 } \beta_{3 } \beta_4 
 +  \frac{1}{4!} \Gamma^{\bar{\beta} \bar{\beta} \bar{\beta} \bar{\beta}}_{ 1 2 3 4}   
 \bar{\beta}_{ -1} \bar{\beta}_{ -2 } \bar{\beta}_{- 3 } \bar{\beta}_{-4}
 \Bigr]. \hspace{7mm}
 \label{eq:S4def}
\end{eqnarray}
The symmetrized quartic vertices are
% \begin{widetext}
 \begin{subequations}
 \begin{eqnarray}
  \Gamma^{ \bar{\beta} \bar{\beta} \beta \beta}_{ 12; 3 4}    & = &
 \sum_{\lambda} \Bigl[
  F_{1+2, \lambda} \Gamma^{\bar{\beta} \bar{\beta}}_{-1,-2, \lambda}
 \Gamma^{\beta \beta}_{ 3, 4, \lambda}
 +   F_{2+3, \lambda} \Gamma^{\bar{\beta} {\beta}}_{-2, 3, \lambda}
 \Gamma^{\bar{\beta} \beta}_{ -1, 4, \lambda}
% \nonumber
% \\ 
% & + &  
 + F_{3+1, \lambda} \Gamma^{\bar{\beta} {\beta}}_{-1, 3, \lambda}
 \Gamma^{\bar{\beta} \beta}_{ -2, 4, \lambda} \Bigr],
 \\
  \Gamma^{ \bar{\beta} \beta \beta \beta}_{ 1; 2 3 4}    & = &
 \sum_{\lambda} \Bigl[
  F_{1+2, \lambda} \Gamma^{\bar{\beta} {\beta}}_{-1,2, \lambda}
 \Gamma^{\beta \beta}_{ 3, 4, \lambda}
 +  ( 2 \leftrightarrow 3) + (2 \leftrightarrow 4)
 \Bigr] ,
 \\
  \Gamma^{ \bar{\beta} \bar{\beta} \bar{\beta} 
 \beta}_{ 1 2 3; 4 }       & = &
 \sum_{\lambda} \Bigl[
  F_{1+2, \lambda} \Gamma^{\bar{\beta} \bar{\beta}}_{-1,-2, \lambda}
 \Gamma^{\bar{\beta} \beta}_{ -3, 4, \lambda}
 +  ( 2 \leftrightarrow 3) + (1 \leftrightarrow 3)
 \Bigr] ,
 \\
   \Gamma^{\beta \beta \beta \beta}_{ 1 2 3 4}    & = & \sum_{\lambda}
 \Bigl[
  F_{1+2, \lambda} \Gamma^{{\beta} {\beta}}_{1,2, \lambda}
 \Gamma^{{\beta} \beta}_{ 3, 4, \lambda}
 +  ( 2 \leftrightarrow 3) + (2 \leftrightarrow 4)
 \Bigr] ,
 \\
  \Gamma^{\bar{\beta} \bar{\beta} \bar{\beta} \bar{\beta}}_{ 1 2 3 4} 
   & = & \sum_{\lambda}
 \Bigl[
  F_{1+2, \lambda} \Gamma^{\bar{\beta} \bar{\beta}}_{-1,-2, \lambda}
 \Gamma^{\bar{\beta} \bar{\beta}}_{ -3, -4, \lambda}
 +  ( 2 \leftrightarrow 3) + (2 \leftrightarrow 4)
 \Bigr] .
 \end{eqnarray}
 \end{subequations}
\end{widetext}
To leading order in $1/S$, the damping of the magnons
is determined by the $1/S^2$ correction to the normal component
of the magnon self-energy, which can be written as
 \begin{eqnarray}
 \Sigma_2 ( K )  & = & \frac{T}{N} \sum_{ K^{\prime}}
  \Gamma^{\bar{\beta} \bar{\beta} \beta \beta}_{ -K , - K^{\prime}, K^{\prime} , K } G_0 ( K^{\prime} )
 \nonumber
 \\
 & = & 
  \frac{T}{N} \sum_{ K^{\prime} \lambda}
 \Bigl[ | \Gamma^{\bar{\beta} \beta}_{ \bd{k} , \bd{k}^{\prime} ,
 \lambda} |^2 F_0 ( K - K^{\prime} , \lambda )
 \nonumber
 \\
 & & \hspace{8mm} +
| \Gamma^{{\beta} \beta}_{ \bd{k} , \bd{k}^{\prime} ,
 \lambda} |^2 F_0 ( K + K^{\prime} , \lambda )
 \Bigr] G_0 ( K^{\prime} ),
 \hspace{10mm}
 \label{eq:self2}
 \end{eqnarray}
where 
 \begin{equation}
 G_0 ( K ) = \frac{1}{ i \omega - E_{\bd{k}} }
 \end{equation}
is the non-interacting magnon Green function.

\subsection{Magnon spectral function and dynamic structure factor} 

\begin{figure}[t]
\includegraphics[width=80mm]{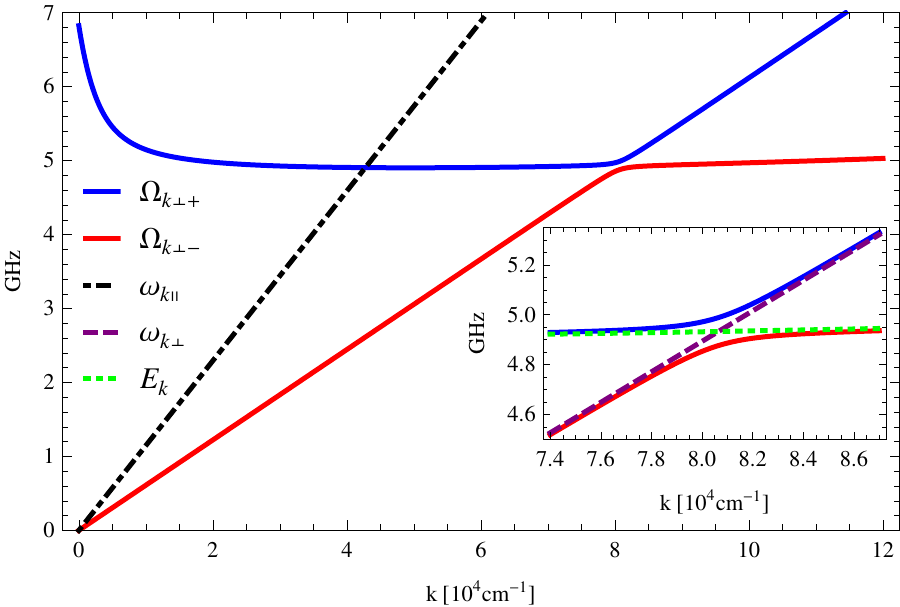}
  \caption{%
(Color online)
Dispersions of the magneto-elastic modes of a thin YIG stripe with
thickness $d = 6.7 \, {\rm \mu m}$ in an external magnetic field $H = 1710 \, {\rm Oe}$, for $\bd{k} = k \bd{e}_z$ parallel to the
in-plane magnetic field. The inset shows a magnified view of the hybridisation at the crossing of magnon and transverse phonon dispersions.
}
  \label{fig:modes}
\end{figure}

Due to the off-diagonal
self-energy $\Pi ( K )$ generated by the magnon-phonon interaction,
the magnon Green function has also an off-diagonal
component, so that we should consider the normal and anomalous
propagators. In terms of the normal irreducible self-energies 
the normal magnon propagator can be written as
 \begin{eqnarray}
 G ( K ) & = &   - T \langle \beta_K \bar{\beta}_K \rangle =
 - \frac{i \omega + E_{\bd{k}} + \Sigma ( - K )}{ D ( K )},
 \hspace{7mm}
 \end{eqnarray}
while the anomalous magnon propagator is
  \begin{eqnarray}
  P ( K ) & = &   - T  \langle \beta_K {\beta}_{-K} \rangle
  =  \frac{ \Pi^{\ast} (  K )}{ D ( K )}.
 \label{eq:pkdef}
 \end{eqnarray}
Here 
 \begin{equation}
 D ( K ) = - [ i \omega - \Sigma_- ( K ) ]^2
 + [ E_{\bd{k}} + \Sigma_+ ( K ) ]^2 - | \Pi ( K ) |^2,
 \end{equation}
can be identified with the determinant of the
inverse matrix Green function, and 
 \begin{equation}
   \Sigma_\pm ( K ) = \frac{1}{2} \left[
 \Sigma ( K ) \pm \Sigma ( - K ) \right].
 \end{equation}
The spectrum of the magneto-elastic modes can be obtained from the roots 
of the equation
 \begin{equation}
 D ( \bd{k} , \omega + i\eta ) =0,
 \end{equation}
with infinitesimal positive $\eta$. Hence, the magneto-elastic modes
are determined by
 \begin{eqnarray}
 & & [ \omega - \Sigma_- ( \bd{k} , \omega ) ]^2
 -  E_{\bd{k}}^2  -  2 E_{\bd{k}} \Sigma_+ ( \bd{k} , \omega )
 \nonumber
 \\ 
 &= &
 [\Sigma_+ ( \bd{k} , \omega )]^2 -  |\Pi ( \bd{k} , \omega )|^2.
 \label{eq:root}
 \end{eqnarray}
For large effective spin $S$ we may
approximate the self-energies by the first order
corrections given in Eqs.~(\ref{eq:sigma1}) and (\ref{eq:pi1}).
Using the explicit polarization basis given
in Eqs.~(\ref{eq:epar})--(\ref{eq:ebot2}) it is easy to see that 
$\Sigma_{-} ( K ) =0$ in this approximation, so that
we may identify $\Sigma_{+} ( K ) = \Sigma ( K )$.
Still, Eq.~(\ref{eq:root}) amounts in general to finding
the solutions of a sixth order
polynomial. However, the last two terms on the
right hand side involving the square of the self-energies
are of order $1/S^2$ and can be neglected; we have checked numerically
that these terms do not have any significant effect for the
parameters relevant to YIG.
The equation for the magneto-elastic modes then reduces to
 \begin{equation}
 \omega^2 - E_{\bd{k}}^2 = 2 E_{\bd{k}} \Sigma_1 ( \bd{k} , \omega ),
 \label{eq:rootsimp}
 \end{equation}
where $\Sigma_1 ( \bd{k} , \omega )$ is defined in Eq.~(\ref{eq:sigma1}).
To further simplify this equation
let us assume that either the energy 
of the longitudinal phonon mode or the energy
of the transverse phonon modes is close to the magnon energy $E_{\bd{k}}$.
In the first case we may approximate
 \begin{equation}
 \Sigma_1 ( \bd{k} , \omega ) \approx 
 \frac{ | \Gamma^{\beta}_{ \bd{k} \parallel} |^2}{ m ( \omega^2 - 
 \omega_{\bd{k} \parallel}^2) },
 \end{equation} 
while in the second case 
\begin{equation}
 \Sigma_1 ( \bd{k} , \omega ) \approx 
 \frac{ | \Gamma^{\beta}_{ \bd{k} \bot} |^2  
    }{ m ( \omega^2 - 
 \omega_{\bd{k} \bot}^2) },
 \end{equation}
where
 \begin{equation}
  | \Gamma^{\beta}_{ \bd{k} \bot} |^2 =  | \Gamma^{\beta}_{ \bd{k} \bot, 1} |^2    +  
  | \Gamma^{\beta}_{ \bd{k} \bot, 2} |^2 .
 \end{equation} 
Eq.~(\ref{eq:rootsimp}) is then bi-quadratic and has the solutions
  \begin{equation}
 \Omega_{\bd{k} \lambda \pm }^2  =
 \frac{ \omega^2_{\bd{k} \lambda} + E^2_{\bd{k}}}{2}
 \pm \sqrt{  \frac{ (\omega^2_{\bd{k} \lambda} - E^2_{\bd{k}})^2 }{4}
 + \Delta_{\bd{k} \lambda}^4 },
 \end{equation}
where
 \begin{eqnarray}
  \Delta_{\bd{k} \lambda}^4 & = &
  2 \frac{E_{\bd{k}}}{m}    | \Gamma^{\beta}_{ \bd{k} \lambda} |^2.
 \end{eqnarray}
The energy dispersion of these modes is shown graphically
in Fig.~\ref{fig:modes} for $\bd{k} = k \bd{e}_z$ parallel to the
in-plane magnetic field.
Note that for this propagation direction
the longitudinal phonon does not hybridise
with the magnon dispersion because
for $k_y =0$ (corresponding to $\theta_{\bd{k}} =0$)
the relevant hybridisation function $\Gamma_{\bd{k} \parallel}$ 
given in Eq.~(\ref{eq:hybparallel}) vanishes.

\begin{figure}[t]
\includegraphics[width=87mm]{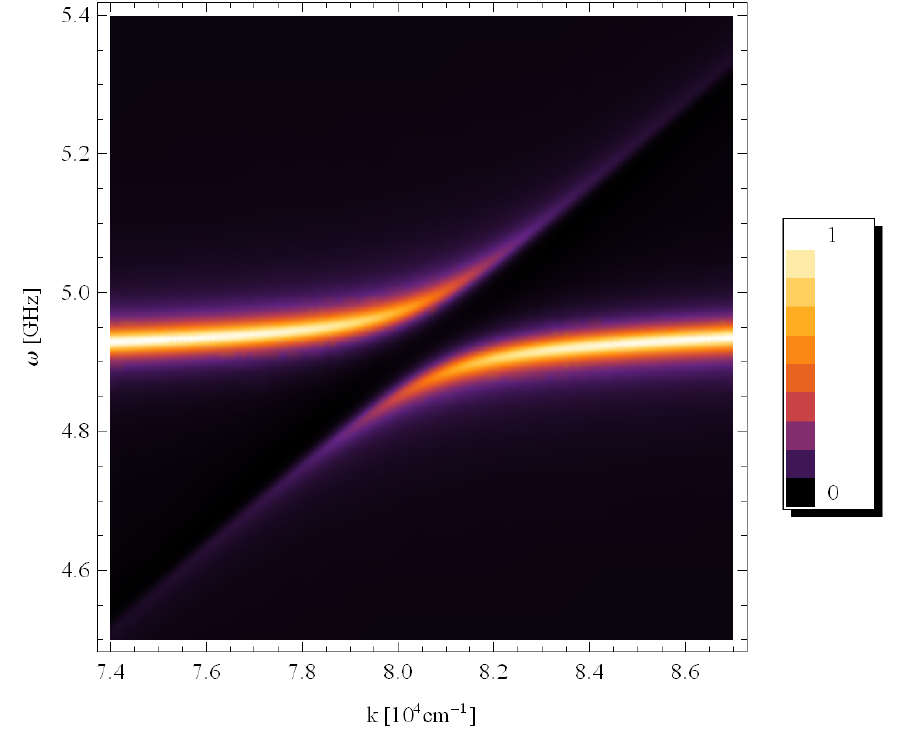}
  \caption{%
(Color online)
Magnon spectral function $ A ( \bd{k} , \omega )$ of a thin YIG stripe with
thickness $d = 6.7 \, {\rm \mu m}$ in an external magnetic field $H = 1710 \, {\rm Oe}$, for $\bd{k} = k \bd{e}_z$ parallel to the
in-plane magnetic field. The Dirac distributions were artificially broadened with a lifetime of $10 \, {\rm ns}$.
}
  \label{fig:spectral}
\end{figure}

\begin{figure}[t]
\includegraphics[width=87mm]{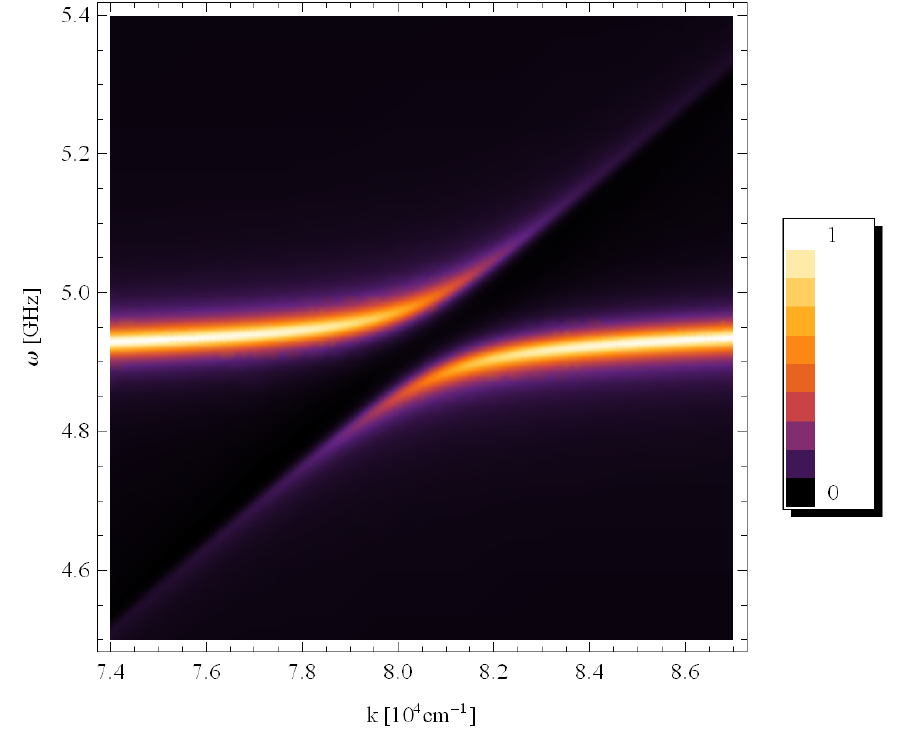}
  \caption{%
(Color online)
Transverse spin dynamic structure factor of a thin YIG stripe at temperature $T = 300 \, {\rm K}$, with
thickness $d = 6.7 \, {\rm \mu m}$ in an external magnetic field $H = 1710 \,  {\rm Oe}$, for $\bd{k} = k \bd{e}_z$ parallel to the
in-plane magnetic field. The Dirac distributions were artificially broadened with a lifetime of $10 \, {\rm ns}$.
}
  \label{fig:dynstruc}
\end{figure}

The normal component of the magnon Green function
is in this approximation
 \begin{eqnarray}
 G ( \bd{k} , i \omega ) & = & \frac{i \omega + E_{\bd{k}} +
 \Sigma_1 ( \bd{k} , i \omega )}{(i \omega)^2 -
 E_{\bd{k}}^2 - 2 E_{\bd{k}} \Sigma_1 ( \bd{k} , i \omega )}.
 \end{eqnarray} 
If $i \omega$ is close to $ \omega_{\bd{k} \lambda}$
this can be approximated by
 \begin{eqnarray}
 G ( \bd{k} , i \omega ) & \approx & 
 \frac{[i \omega + E_{\bd{k}}]
 [ ( i \omega)^2 - \omega_{\bd{k} \lambda}^2 ] 
 + \frac{| \Gamma^{\beta}_{\bd{k} \lambda} |^2}{m} }{
 [ (i \omega)^2 - E_{\bd{k}}^2]
 [( i \omega)^2 - \omega_{\bd{k} \lambda}^2 ]
   - 2 E_{\bd{k}} \frac{| \Gamma^{\beta}_{ \bd{k} \lambda} |^2}{m}   }.
 \hspace{10mm}
 \end{eqnarray} 
After analytic continuation to the real frequency axis
we obtain for the corresponding spectral function
 \begin{eqnarray}
 & & A ( \bd{k} , \omega )  =  - \frac{1}{\pi} {\rm Im} 
 G ( \bd{k} , \omega + i \eta ) 
 \nonumber
 \\
 & \approx & Z_{\bd{k} \lambda}  (\omega ) \sum_{ s = \pm }
 s [ \delta ( \omega - \Omega_{ \bd{k} \lambda s } )
 +  \delta ( \omega + \Omega_{ \bd{k} \lambda s } )  ],
 \hspace{7mm}
 \end{eqnarray}
where
\vspace*{-.3cm}
  \begin{equation}
 Z_{\bd{k} \lambda}  (\omega )  =  \frac{ 
[\omega + E_{\bd{k}}]
 [  \omega^2 - \omega_{\bd{k} \lambda}^2 ] 
 + \frac{| \Gamma^{\beta}_{\bd{k} \lambda} |^2}{m} }{
 2 \omega [ \Omega^2_{ \bd{k} \lambda + }
 -  \Omega^2_{ \bd{k} \lambda - } ] }.
 \end{equation}
An intensity  plot of the magnon spectral function for YIG is
shown in Fig.~\ref{fig:spectral}.
One clearly sees the transfer of spectral weight 
 between the magnon and the phonon branch at the crossing point.
Actually, the Brillouin light scattering intensity is proportional
to the transverse spin structure factor \cite{Jorzick99,Cottam86}
 \begin{equation}
 S_{\bot} ( \bd{k} , \omega ) = \int_{ - \infty}^{\infty} 
 \frac{dt}{2 \pi} e^{ i \omega t} \langle
 S^x_{ - \bd{k} } (0 ) S^{x}_{\bd{k}} ( t ) 
 + S^y_{ - \bd{k} } (0 ) S^{y}_{\bd{k}} ( t )  \rangle,
 \end{equation}
where the Fourier transform of the spin-operator
is defined in Eq.~(\ref{eq:FTspin}).
To leading order in spin wave theory we obtain
\vspace*{-.2cm}
 \begin{eqnarray}
S_{\bot} ( \bd{k} , \omega ) & = &
  \frac{S}{ 1 - e^{ - \omega /T }}
			 \nonumber
 \\
  & &
	\times
\Bigl\{ ( u_{\bd{k}}^2 + | v_{\bd{k}} |^2 ) 
 \left[ A ( - \bd{k} , \omega ) - A ( \bd{k} , - \omega ) 
 \right]
		 \nonumber
 \\
  & &
-2 \left( u_{\bd{k}} v^{\ast}_{\bd{k}} 
 + u_{\bd{k}} v_{\bd{k}} \right) B ( \bd{k} , \omega )
 \Bigr\},
 \nonumber
 \\
 & &
 \end{eqnarray}
where 
\vspace*{-.3cm}
 \begin{equation}
 B ( \bd{k} , \omega ) =   - \frac{1}{\pi} {\rm Im} 
 P ( \bd{k} , \omega + i \eta ) 
 \end{equation}
is the spectral function of the anomalous magnon Green function
$P ( \bd{k} , i \omega)$, which for imaginary frequencies is defined by
Eq.~(\ref{eq:pkdef}).
We obtain
 \begin{eqnarray}
 B ( \bd{k} , \omega )  
 & \approx & Y_{\bd{k} \lambda}  (\omega ) \sum_{ s = \pm }
 s [ \delta ( \omega - \Omega_{ \bd{k} \lambda s } )
 +  \delta ( \omega + \Omega_{ \bd{k} \lambda s } )  ],
 \nonumber
 \\
 & &
 \end{eqnarray}
where
  \begin{equation}
 Y_{\bd{k} \lambda}  (\omega )  =  - \frac{ 
   \Gamma^{\bar{\beta}}_{-\bd{k} \bot, 1}    
   \Gamma^{\bar{\beta}}_{\bd{k} \bot, 1}   
   +
 \Gamma^{\bar{\beta}}_{-\bd{k} \bot, 2}    
   \Gamma^{\bar{\beta}}_{\bd{k} \bot, 2}   
    }{
 2 m\omega [ \Omega^2_{ \bd{k} \lambda + }
 -  \Omega^2_{ \bd{k} \lambda - } ] }.
 \end{equation}
An intensity plot of the transverse dynamic structure factor is
shown in Fig.~\ref{fig:dynstruc}.
Note that the qualitative behavior is very similar to the
behavior of the magnon spectral function, which is due to the fact
that the Bogoliubov transformation has only a small effect because of the smallness
of the Boguliubov coefficient $v_{\bd{k}}$ in the entire Brillouin zone.
\vspace{2mm}

\section{Magnon damping}
\label{sec:damp}
 
\begin{figure}[t]
\includegraphics[width=80mm]{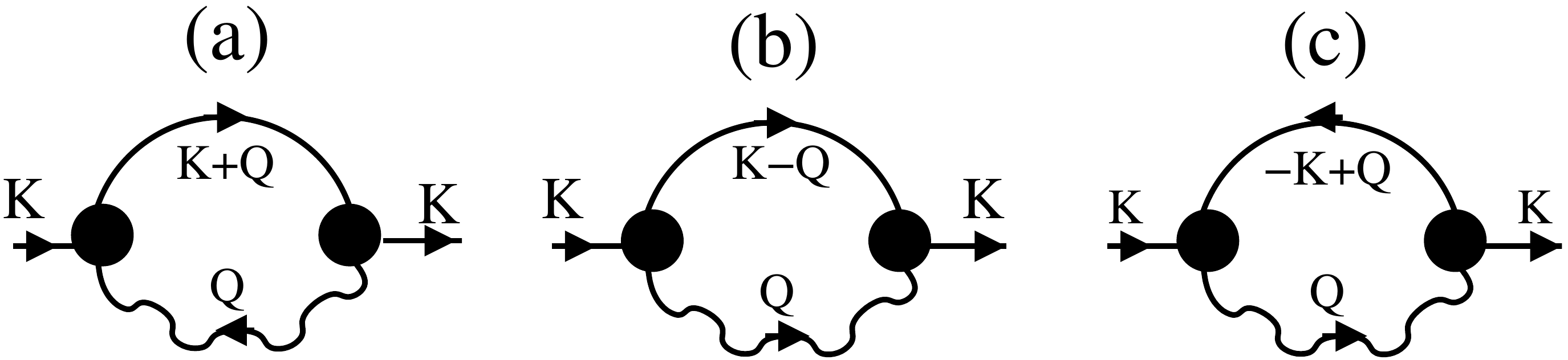}
  \caption{%
%(Color online)
These Feynman diagrams gives rise to
the three contributions to the 
magnon decay rate given in Eqs.~(\ref{eq:chera} -- \ref{eq:dampconf}):
(a) Cherenkov process with absorption of a virtual phonon;
(b)  Cherenkov process with emission of a virtual phonon;
(c) confluent process.
The solid arrows represent magnon propagators while the wavy arrows 
represent phonon propagators. The arrows represent the direction
of the energy-momentum flow. Black dots represent the relevant components 
of the  three-legged vertices  $\bd{\Gamma}^{\bar{\beta} \beta}$ and 
$\bd{\Gamma}^{\beta \beta}$ defined in Eqs.~(\ref{eq:Gammabb1}) and
(\ref{eq:Gammabb2}).
}
  \label{fig:feynman}
\end{figure}

In this section we calculate the decay rate (i.e., the damping) of the magnons
due to the magnon-phonon interaction in YIG at room temperature.
The damping $\gamma ( {\bd{k}} )$ of magnons with wave vector $\bd{k}$
and energy $E_{\bd{k}}$
can be obtained from the
imaginary part of the self-energy $\Sigma ( K ) = \Sigma ( \bd{k} , i \omega )$
after analytic continuation to the real frequency axis,
 \begin{equation}
 \gamma ( \bd{k} ) = - {\rm Im}  \Sigma ( \bd{k} , i \omega 
 \rightarrow E_{\bd{k}} + i \eta ).
 \end{equation}
To leading order in $1/S$ the damping is determined by the
second order self-energy  given in Eq.~(\ref{eq:self2});
after carrying out the frequency sum we obtain
 \begin{widetext}
 \begin{eqnarray}
 \Sigma_2 ( \bd{k} , i \omega ) & = & \frac{1}{N} \sum_{\bd{k}^{\prime} 
 \lambda}  \Biggl\{ 
 \frac{  | \Gamma^{\bar{\beta} \beta}_{ \bd{k} , \bd{k}^{\prime} ,
 \lambda} |^2}{ 2 m  \omega_{ \bd{k} - \bd{k}^{\prime}  \lambda}}
  \left[ 
 \frac{ b ( \omega_{\bd{k} - \bd{k}^{\prime} \lambda} ) - b ( 
 E_{\bd{k}^{\prime} } ) }{ i \omega +   
 \omega_{\bd{k} - \bd{k}^{\prime} \lambda} - E_{\bd{k}^{\prime}} }
 + \frac{ 1+ b ( \omega_{\bd{k} - \bd{k}^{\prime} \lambda} ) + b ( 
 E_{\bd{k}^{\prime} } ) }{ i \omega -   
 \omega_{\bd{k} - \bd{k}^{\prime} \lambda} - E_{\bd{k}^{\prime}} }
 \right]
 \nonumber
 \\
 & & \hspace{10mm} -
\frac{  | \Gamma^{{\beta} \beta}_{ \bd{k} , \bd{k}^{\prime} ,
 \lambda} |^2}{ 2 m \omega_{ \bd{k} + \bd{k}^{\prime}  \lambda}}
  \left[ 
 \frac{ 1+ b ( \omega_{\bd{k} + \bd{k}^{\prime} \lambda} ) + b ( 
 E_{\bd{k}^{\prime} } ) }{ i \omega +   
 \omega_{\bd{k} + \bd{k}^{\prime} \lambda} + E_{\bd{k}^{\prime}} }
 + \frac{  b ( \omega_{\bd{k} + \bd{k}^{\prime} \lambda} ) - b ( 
 E_{\bd{k}^{\prime} } ) }{ i \omega -   
 \omega_{\bd{k} + \bd{k}^{\prime} \lambda} + E_{\bd{k}^{\prime}} }
 \right]
 \Biggr\}.
 \end{eqnarray}
Here $b ( \omega ) = 1/( e^{\omega / T } -1 )$ is the Bose function.
The corresponding damping function off resonance is

 \begin{eqnarray}
 \gamma_2 ( \bd{k} , \omega )  & = &  - {\rm Im}  \Sigma_2 ( \bd{k} , \omega + i \eta )
 \nonumber
 \\
 & = &   ( 1 - e^{- \omega / T } ) \frac{\pi}{ N}
 \sum_{\bd{k}^{\prime} 
 \lambda}  \Biggl\{ 
 \frac{  | \Gamma^{\bar{\beta} \beta}_{ \bd{k} , \bd{k}^{\prime} ,
 \lambda} |^2}{ 2 m \omega_{ \bd{k} - \bd{k}^{\prime}  \lambda}}
 \Bigl[
 \delta ( \omega - E_{\bd{k}^{\prime}}+ 
 \omega_{ \bd{k} - \bd{k}^{\prime}  \lambda} )
 ( 1 + b ( E_{\bd{k}^{\prime}} ) ) 
 b ( \omega_{ \bd{k} - \bd{k}^{\prime}  \lambda} )
 \nonumber
 \\
 & & \hspace{45mm} + \delta ( \omega - E_{\bd{k}^{\prime}}- 
 \omega_{ \bd{k} - \bd{k}^{\prime}  \lambda} )
 ( 1 + b ( E_{\bd{k}^{\prime}} ) ) 
 (1 + b ( \omega_{ \bd{k} - \bd{k}^{\prime}  \lambda} ))
 \Bigr]
 \nonumber
 \\
  & & \hspace{28mm}  + \frac{  | \Gamma^{{\beta} \beta}_{ \bd{k} , \bd{k}^{\prime} ,
 \lambda} |^2}{ 2 m \omega_{ \bd{k} + \bd{k}^{\prime}  \lambda}}
 \Bigl[
 \delta ( \omega + E_{\bd{k}^{\prime}}+ 
 \omega_{ \bd{k} + \bd{k}^{\prime}  \lambda} )
  b ( E_{\bd{k}^{\prime}} )  
 b ( \omega_{ \bd{k} + \bd{k}^{\prime}  \lambda} )
 \nonumber
 \\
 & & \hspace{45mm} + \delta ( \omega + E_{\bd{k}^{\prime}}- 
 \omega_{ \bd{k} - \bd{k}^{\prime}  \lambda} )
 b ( E_{\bd{k}^{\prime}} )  
 (1 +  b ( \omega_{ \bd{k} + \bd{k}^{\prime}  \lambda} ))
 \Bigr]
\Biggr\}.
 \end{eqnarray}
 \end{widetext}
Since the experiments of interest to us are performed at room
temperature which is large compared with all
other energy scales, we may use the high temperature expansion
of the Bose functions, $b ( \omega ) \approx T / \omega$.
Setting now $\omega = E_{\bd{k}}$ we obtain for the magnon damping on
resonance at high temperatures,
 \begin{eqnarray}
 \gamma_2 ( \bd{k} ) & = & \frac{ \pi T  E_{\bd{k}}  }{2m N} 
 \sum_{\bd{k}^{\prime} \lambda} 
 \Biggl\{ 
 \frac{   | \Gamma^{\bar{\beta} \beta}_{ \bd{k} , \bd{k}^{\prime} ,
 \lambda} |^2  }{   E_{\bd{k}^{\prime}}
 \omega^2_{ \bd{k} - \bd{k}^{\prime}  \lambda}}
 \Bigl[
 \delta ( E_{\bd{k}} - E_{\bd{k}^{\prime}}+ 
 \omega_{ \bd{k} - \bd{k}^{\prime}  \lambda} )
 \nonumber
 \\
 & & \hspace{35mm}
 + \delta ( E_{\bd{k}} - E_{\bd{k}^{\prime}}- 
 \omega_{ \bd{k} - \bd{k}^{\prime}  \lambda} )
 \Bigr]
 \nonumber
 \\
 &  & \hspace{15mm} +
  \frac{   | \Gamma^{{\beta} \beta}_{ \bd{k} , \bd{k}^{\prime} ,
 \lambda} |^2  }{  E_{\bd{k}^{\prime}}
 \omega^2_{ \bd{k} + \bd{k}^{\prime}  \lambda}}
 \delta ( E_{\bd{k}} + E_{\bd{k}^{\prime}}- 
 \omega_{ \bd{k} + \bd{k}^{\prime}  \lambda} )
 \Biggr\}
 \nonumber
 \\
 & =  &   \gamma_{2a}^{\rm Che} ( \bd{k} ) +  \gamma_{2b}^{\rm Che} ( \bd{k} )
 +\gamma_2^{\rm con} ( \bd{k} ),
 \label{eq:damp2}
 \end{eqnarray}
where 
 \begin{eqnarray}
   \gamma_{2a}^{\rm Che} ( \bd{k} ) 
 & =  &
\frac{ \pi T  E_{\bd{k}}  }{2m N} 
 \sum_{\bd{q} \lambda} 
 \frac{   | \Gamma^{\bar{\beta} \beta}_{ \bd{k} , \bd{k}+ \bd{q} ,
 \lambda} |^2  }{  E_{\bd{k} + \bd{q}}  
 \omega^2_{ \bd{q}  \lambda}}
 \delta ( E_{\bd{k}} - E_{\bd{k} + \bd{q}}+ 
 \omega_{ \bd{q}  \lambda} ),
 \nonumber
 \\
 & &
 \label{eq:chera}
 \\
 \gamma_{2b}^{\rm Che} ( \bd{k} ) 
    & = & 
\frac{ \pi T  E_{\bd{k}}  }{2m N} 
 \sum_{\bd{q} \lambda} 
 \frac{   | \Gamma^{\bar{\beta} \beta}_{ \bd{k} , \bd{k}- \bd{q} ,
 \lambda} |^2  }{  E_{\bd{k} - \bd{q}}  
 \omega^2_{ \bd{q}  \lambda}}
 \delta ( E_{\bd{k}} - E_{\bd{k} - \bd{q}}- 
 \omega_{ \bd{q}  \lambda} ),
 \nonumber
 \\
 & &
 \label{eq:cherb}
 \\
\gamma_2^{\rm con} ( \bd{k} ) 
 & =  &
\frac{ \pi T  E_{\bd{k}}  }{2m N} 
 \sum_{\bd{q} \lambda} 
  \frac{   | \Gamma^{{\beta} \beta}_{ \bd{k} , - \bd{k} + \bd{q} ,
 \lambda} |^2  }{ E_{- \bd{k} + \bd{q} }
 \omega^2_{ \bd{q}  \lambda}}
 \delta ( E_{\bd{k}} + E_{  - \bd{k} + \bd{q}}- 
 \omega_{ \bd{q}   \lambda} ).
 \nonumber
 \\
 & &
 \label{eq:dampconf}
 \end{eqnarray}
The contributions   $\gamma_{2a}^{\rm Che} ( \bd{k} ) $
and $\gamma_{2b}^{\rm Che} ( \bd{k} ) $ are due to the Cherenkov type 
process where a magnon
with energy $E_{\bd{k}}$  emits or absorbs
a phonon with energy $\omega_{\bd{q}}$ and decays into a magnon with
energy $E_{\bd{k} \pm \bd{q}}$.
The last contribution
$\gamma_2^{\rm con} ( \bd{k} ) $ 
describes a confluent scattering process where
two magnons with energies
$E_{\bd{k}} $ and $E_{\bd{-k} + \bd{q}}$ decay into a phonon
with energy $\omega_{\bd{q}}$.
The corresponding Feynman diagrams are shown in Fig.~\ref{fig:feynman}.
Taking into account that for YIG 
the interaction vertices in the quasi-particle basis can be approximated
by the corresponding interaction vertices in the Holstein-Primakoff
basis, we may approximate the squared matrix elements in the 
above expressions by
 \begin{eqnarray}
  | \Gamma^{\bar{\beta} \beta}_{ \bd{k} , \bd{k} \pm \bd{q} ,
 \lambda} |^2 & \equiv & | \bd{e}_{\bd{q} \lambda}^{\ast} 
 \cdot  \bd{\Gamma}^{\bar{\beta} \beta}_{ \bd{k} , \bd{k} \pm \bd{q} } |^2
  \approx  | \bd{e}_{\bd{q} \lambda}^{\ast} 
 \cdot  \bd{\Gamma}^{\bar{b} b}_{ \bd{k} , \bd{k} \pm \bd{q} } |^2
 \nonumber
 \\
 & = & 
   | \bd{e}_{\bd{q} \lambda}^{\ast}  \cdot  \bd{U}_{ \pm \bd{q}} |^2
  \equiv  | \bd{q} |^2 {U}^2_{\lambda} ( \hat{\bd{q}} ),
 \end{eqnarray}
 \begin{eqnarray}
  | \Gamma^{{\beta} \beta}_{ \bd{k} , -\bd{k}+ \bd{q} ,
 \lambda} |^2 & \equiv & | \bd{e}_{\bd{q} \lambda}^{\ast} 
 \cdot  \bd{\Gamma}^{{\beta} \beta}_{ \bd{k} , -\bd{k}+ \bd{q} } |^2
  \approx  | \bd{e}_{\bd{q} \lambda}^{\ast} 
 \cdot  \bd{\Gamma}^{{b} b}_{ \bd{k} , -\bd{k}+ \bd{q} } |^2
 \nonumber 
 \\
 & = & 
   | \bd{e}_{\bd{q} \lambda}^{\ast}  \cdot  \bd{V}_{\bd{q}} |^2
  \equiv  | \bd{q} |^2 {V}^2_{\lambda} ( \hat{\bd{q}} ),
 \end{eqnarray}
where
 \begin{eqnarray}
{U}^2_{\lambda} ( \hat{\bd{q}} ) & = & \frac{ B_{\parallel}^2}{S^2}
 \left|  \bd{e}_{\bd{q} \lambda}^{\ast}  \cdot ( \hat{q}_x 
 \bd{e}_y + \hat{q}_y \bd{e}_x - 2 \hat{q}_z \bd{e}_z ) \right|^2,
 \\
{V}^2_{\lambda} ( \hat{\bd{q}} ) & = &
 \frac{ B_{\parallel}^2}{S^2}
 \left|  \bd{e}_{\bd{q} \lambda}^{\ast}  \cdot ( \hat{q}_x 
 \bd{e}_x - \hat{q}_y \bd{e}_y ) \right|^2
 \nonumber
 \\
 & + & 
 \frac{ B_{\bot}^2}{S^2}
 \left|  \bd{e}_{\bd{q} \lambda}^{\ast}  \cdot ( \hat{q}_x 
 \bd{e}_y + \hat{q}_y \bd{e}_x ) \right|^2.
 \end{eqnarray}
Here $\hat{q}_{\alpha} = q_{\alpha} / | \bd{q} |$ are 
the components of the unit vector in the direction of $\bd{q}$.
Using the phonon basis in Eqs.~(\ref{eq:epar})--(\ref{eq:ebot2})
we obtain for $q_x=0$,
 \begin{subequations}
 \begin{eqnarray}
 U^{2}_{\parallel} ( \hat{\bd{q}} ) & = & \frac{ B_{\parallel}^2}{S^2} (
 \hat{q}_y^2 - 2 \hat{q}_z^2 )^2 =   \frac{ B_{\parallel}^2}{S^2}
 ( 1 - 3 \cos^2 \theta_{\bd{q}} )^2,
 \hspace{7mm}
 \\
 U^{2}_{\bot 1} ( \hat{\bd{q}} ) & = & \frac{ B_{\parallel}^2}{S^2} 
 ( 3 \hat{q}_y  \hat{q}_z )^2  =   \frac{ B_{\parallel}^2}{S^2} \frac{9}{4}
 \sin^2 (2 \theta_{\bd{q}} ),
 \\
 U^{2}_{\bot 2} ( \hat{\bd{q}} ) & = & 0,
 \end{eqnarray}
 \end{subequations}
and
 \begin{subequations}
 \begin{eqnarray}
 V^{2}_{\parallel} ( \hat{\bd{q}} ) & = & \frac{ B_{\parallel}^2}{S^2}
\hat{q}_y^4 =   \frac{ B_{\parallel}^2}{S^2}
\sin^4 \theta_{\bd{q}},
 \\
 V^{2}_{\bot 1} ( \hat{\bd{q}} ) & = & \frac{ B_{\parallel}^2}{S^2} 
 (  \hat{q}_y  \hat{q}_z )^2  =   \frac{ B_{\parallel}^2}{S^2} \frac{1}{4}
 \sin^2 (2 \theta_{\bd{q}} ),
 \\
 V^{2}_{\bot 2} ( \hat{\bd{q}} ) & = & \frac{ B_{\bot}^2}{S^2} 
\hat{q}_y^2   =   \frac{ B_{\bot}^2}{S^2}
 \sin^2  \theta_{\bd{q}} ,
 \end{eqnarray}
 \end{subequations}
where we have set $\hat{q}_z = \cos \theta_{\bd{q}}$ and
$\hat{q}_y = \sin \theta_{\bd{q}}$.

\subsection{Dipolar regime: Theory}

\begin{figure}[t]
\includegraphics[width=70mm]{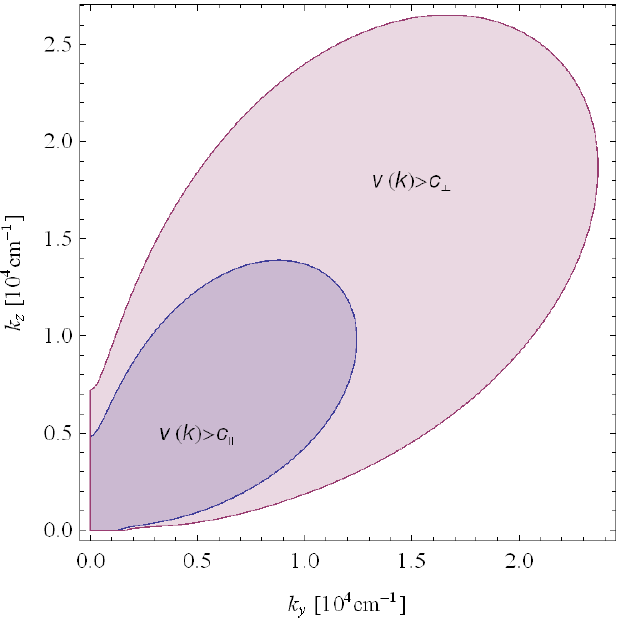}
  \caption{%
(Color online)
The shaded areas denote the momentum range
where the group velocity $v ( \bd{k} )$ of the magnons exceeds
either the longitudinal or the transverse
phonon velocity. The parameters are the same as in Fig.~\ref{fig:dynstruc}.
}
\label{fig:velocities}
\end{figure}

\begin{figure}[t]
\includegraphics[width=80mm]{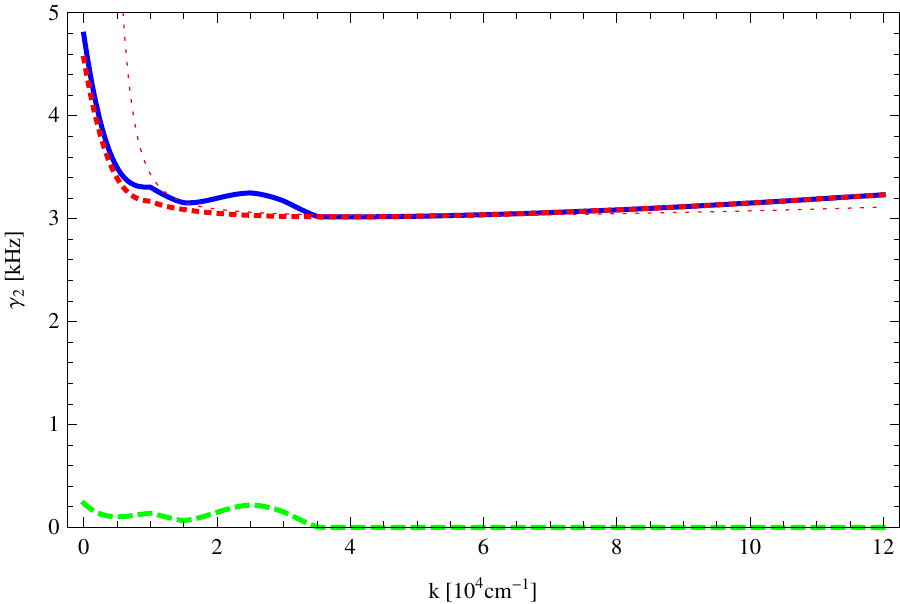}
  \caption{%
(Color online)
Numerical evaluation of our result
(\ref{eq:damp2}) for the damping rate of magnons in a thin YIG
stripe at temperature $T = 300 \, {\rm K}$, in the dipolar momentum regime.
The plot is for a thin stripe of thickness $d = 6.7 {\rm \mu m}$ in an 
external magnetic field $H = 1710 \, {\rm Oe}$, for $\bd{k} = k \bd{e}_z$ 
parallel to the in-plane magnetic field.
Solid lines correspond to the total damping rate, while
the dashed and the dotted lines are the
contribution from the Cherenkov  and the confluent processes, respectively.
The corresponding thin dotted line is the approximation 
(\ref{eq:dampconf2}) in the dipolar momentum regime.
}
\label{fig:dampdipolar}
\end{figure}

In the long-wavelength regime $ | \bd{k} | \lesssim \sqrt{ \Delta / \rho_s }$ the
behavior of the magnon damping (\ref{eq:damp2}) strongly
depends on the size
$v ( \bd{k} ) = | \bd{v} ( \bd{k} ) |$
of the group velocity $\bd{v} ( \bd{k} ) = \nabla_{\bd{k}} E_{\bd{k}}$
of the magnons in comparison with the phonon velocities.
In Fig.~\ref{fig:velocities} we show the momentum range where the
magnon velocity
$v ( \bd{k} )$ exceeds the phonon velocities $c_{\lambda}$.
In regime around the minima
of the dispersion, the velocity $v ( \bd{k} )$ is
small compared with the phonon velocities, while
at very small wave vectors $ v ( \bd{k} ) > c_{\lambda}$. 
In this regime around the minima of the dispersion
the decay rate the magnons is dominated by the
confluent process given in Eq.~(\ref{eq:dampconf}) because 
the Cherenkov processes are kinematically suppressed.
In fact, in a substantial regime around
dispersion minima the quasi-particle velocity is small
compared with the phonon velocities, so that we may approximate
$E_{- \bd{k} + \bd{q}} \approx E_{\bd{k}} - \bd{v} ( \bd{k} )
 \cdot \bd{q}$ and expand the decay rate 
in powers of $v ( \bd{k} ) / c_{\lambda}$.
The momentum integration in Eq.~(\ref{eq:dampconf}) can then be
carried out and we obtain
 \begin{eqnarray}
 \gamma_2^{\rm con} ( \bd{k} ) & = & \frac{ T E_{\bd{k}}}{4 S^2}
	\frac{a^2}{m} \Biggl[ \frac{ B_{\bot}^2}{ c_{\bot}^4}
 \left( 1 + \frac{3}{4} \frac{ v_{y}^2( \bd{k} )}{c_{\bot}^2} +
  \frac{ v_{z}^2( \bd{k} )}{4c_{\bot}^2}  \right)
 \nonumber
 \\
 & &\hspace{12mm}  +\frac{ B_{\parallel}^2}{ 4 c_{\bot}^4}
 \left( 1 +  \frac{ v_{y}^2( \bd{k} )}{2c_{\bot}^2} +
  \frac{ v_{z}^2( \bd{k} )}{2c_{\bot}^2}  \right)
 \nonumber
 \\
 & &\hspace{12mm}  + \frac{3}{4} \frac{ B_{\parallel}^2}{  c_{\parallel}^4}
 \left( 1 +  \frac{5}{6} \frac{ v_{y}^2( \bd{k} )}{c_{\parallel}^2} +
  \frac{ v_{z}^2( \bd{k} )}{6c_{\parallel}^2}  \right)
 \Biggr]. 
\nonumber
\\
 \label{eq:dampconf2}
 \hspace{7mm}
 \end{eqnarray}
The confluent contribution to the
high-temperature damping rate in the dipolar regime
is shown graphically as the thin dotted line in Fig.~\ref{fig:dampdipolar}.
On the other hand, 
the Cherenkov-type process contributes
only for very small $\bd{k}$, when $v ( \bd{k} ) \gtrsim c_{\lambda}$, see Fig.~\ref{fig:velocities}.
However, for those momenta it is no longer valid to linearize the dispersion 
due to the strong effect of the Bogoliubov transformation, and no analytical 
approximation can be obtained. Therefore we only present the numerical solution
as dashed line in Fig.~\ref{fig:dampdipolar}.
Obviously,  apart from an enhancement for  $k \lesssim 2 \times 
10^{4} \, {\rm cm}^{-1}$
the damping exhibits a rather weak dependence 
on the wave vector in this regime.

\subsection{Dipolar regime: Experiment}

For a comparison of our calculation with experiments
one should keep in mind that we have only considered the
contribution from the magnon-phonon interactions on the 
damping of the magnons. Of course, 
in the real system there are other sources leading to
magnon decay, such as 
magnon-magnon interactions or the elastic scattering of
magnons by impurities. We therefore expect that 
the magnon damping due to magnon-phonon interactions is a lower limit
to the experimentally observed magnon damping rate.
In fact, our experimental data presented below
show that in the dipolar regime
magnon-phonon interactions are not the dominant
source of magnon damping.
\begin{figure}[t]
\includegraphics[width=80mm]{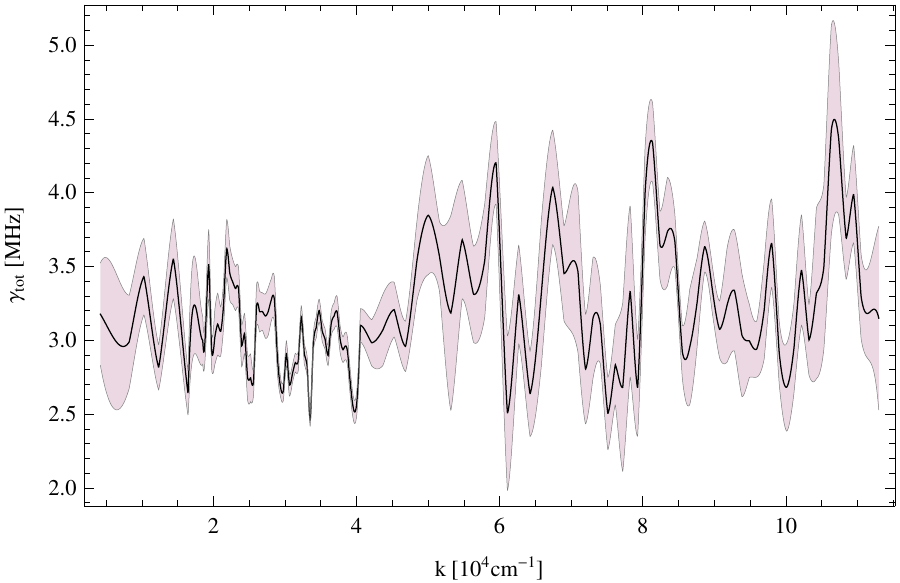}
  \caption{%
Experimentally determined dependence of the total magnon damping 
$\gamma_{\rm tot}$ on the in-plane wave vector in the dipolar 
regime where the magnon dispersion
is dominated by the competition between the dipole-dipole and the
exchange interaction.
The shaded regions represent the estimated experimental uncertainties.
}
\label{Figure_1}
\end{figure}

In order to determine the relaxation time of different groups of magnons, 
a measurements of the spectral distribution of magnon gas densities 
as a function of the frequency and wave vector 
using time- and wave-vector-resolved BLS spectroscopy \cite{Sandweg} 
has been performed. 
Due to technical reasons only 
in-plane wave vectors  from $0$ up to the $k_z^{\rm max} = 11 \times 10^4\,{\rm cm}^{-1}$ are accessible within our apparatus.
The measurements were performed using a YIG film
with thickness $6.7\,\mu$m, which was 
liquid-phase epitaxialy grown on a $500\,\mu$m thick gallium gadolinium garnet substrate. 

The magnon spectrum was populated by intensive thermalization \cite{DemokritovNature, Hick12, Demokritov} of magnons, 
which were injected by the parallel parametric pumping technique \cite{Serga} at half of the pumping frequency 
$f_p = 13.62\,$GHz. The bias magnetic field of $H = 1710\,$Oe was tuned to provide the excitation of 
parametric magnons at the ferromagnetic resonance frequency. In this case the direct 
transition of magnons to the bottom of spin wave spectrum is prohibited by conservation laws. This ensures high efficiency of multi-stage four-magnon scattering which is necessary for thermalization and thus population of the spectrum. 

We have measured the redistribution of thermalized magnons along the fundamental 
backward-volume magnetostatic spin wave
mode as a  function of time and wave vector. 
After switching off the pumping, the magnons are allowed to relax freely.
By fitting the relaxation times for different groups of 
thermalized magnons we were able to extract the damping rates. 
The obtained dependence of measured total damping rate $\gamma_{\rm tot}$ 
on the in-plane wave vector for dipolar-exchange spin waves 
is shown in Fig.~\ref{Figure_1}. 
Obviously,
the value of relaxation rate is roughly three orders of magnitude
larger than  our calculated  relaxation rate due to magnon-phonon interactions
shown  Fig.~\ref{fig:dampdipolar}. We thus conclude that in the long-wavelength
dipolar regime other relaxation channels 
(in particular two-magnon scattering 
processes\cite{Sparks, ChumakPRB, Patton, ChumakAPL})
dominate the magnon damping.
The rather irregular behavior of the measured damping rate in Fig.~\ref{Figure_1}
suggests that elastic scattering of magnons by impurities might play
an important role in this regime.
Note that within the tolerance limits of the experiment the measured
relaxation rate has a  rather weak dependence on the in-plane wave vectors
in the entire accessible range of wave vectors.
In this respect the experimental results agree with our
prediction of a momentum-independent damping rate in this regime.

Unfortunately, microscopic calculations of the magnon decay rates
at room temperature, taking magnon-impurity and
magnon-magnon interactions into account, are not available 
 in the momentum range relevant for
our experiment.
One should keep in mind, however,
that magnon-impurity  scattering can only explain the momentum-relaxation  of the 
magnon gas; for the equilibration of the different temperatures of the 
magnon and the phonon systems
magnon-phonon interactions are essential.

%Nevertheless, this measurements are consistent with presented theory in the following way. Measured relaxation rate has practically non-dependant behavior on in-plane wave number within the tolerance limits of the measurement in all accessible range of wave numbers. The two-magnon scattering processes due to it's nature is practically independent on wave number for this region of spectrum. Contributions from the Cherenkov and the confluent processes in this case are significantly lower and due to it's flat shape (see Fig.~\ref{fig:dampdipolar}) does not change the flatness of the resulting relaxation rate figure.

%It is necessary to note that two-magnon scattering contributes to thermalization of a magnon gas in a finite wave vector range but does not lead to equilibration of magnon and phonon baths temperatures.

\subsection{Exchange regime}

\begin{figure}[t]
\includegraphics[width=80mm]{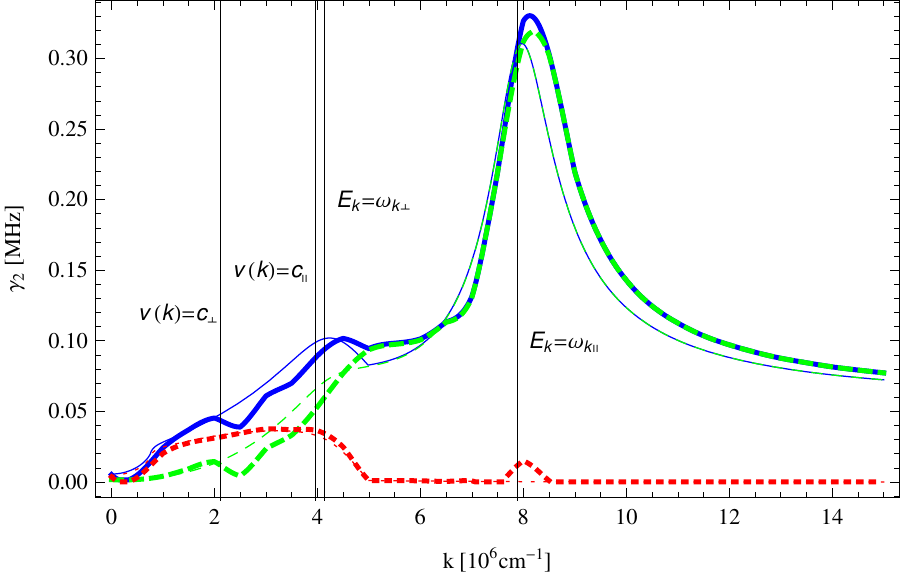}  
  \caption{%
(Color online)
Numerical evaluation of our result
(\ref{eq:damp2}) for the damping rate of magnons in a thin YIG
stripe at temperature $T = 300 \, {\rm K}$, in the exchange momentum regime.
The plot is for a thin stripe of thickness $d = 6.7 \, {\rm \mu m}$ in an external magnetic field $H = 1710 \, {\rm Oe}$, for $\bd{k} = k \bd{e}_z$ parallel to the
in-plane magnetic field. Solid lines correspond to the total damping rate, while
the dashed and the dotted lines are the
contribution from the Cherenkov  and the confluent processes, respectively.
The corresponding thin lines are the approximations 
in the exchange momentum regime, see
Eqs.~(\ref{eq:gammachex}) and (\ref{eq:gammaconex}).
}
  \label{fig:damp}
\end{figure}

For wave vectors in the regime where the exchange energy
 $ \rho_s \bd{k}^2$  exceeds the characteristic dipolar energy
$\Delta$
 we may ignore the dipole-dipole
interaction in the  magnon dispersion  (\ref{eq:Eklong})
and approximate the long-wavelength magnon dispersion by
$E_{\bd{k}} \approx h + \rho_s \bd{k}^2$.
Then the evaluation of the integrals in
 Eqs.~(\ref{eq:chera})--(\ref{eq:dampconf}) simplifies.
We obtain for the Cherenkov part 
 \begin{eqnarray}
\gamma_{2}^{\rm Che} ( \bd{k} )  & \equiv &
 \gamma_{2a}^{\rm Che} ( \bd{k} ) +
 \gamma_{2b}^{\rm Che} ( \bd{k} )
  \nonumber
 \\
 & = &    \frac{T E_{\bd{k}} }{2 } \frac{m_s}{m}
 \sum_{\lambda} \frac{a^2}{c_{\lambda}^2}
\nonumber
\\
& &\times
 \int_0^{2 \pi} \frac{ d \varphi}{2 \pi}
 \frac{ U_{\lambda}^2 ( \hat{\bd{q}}_\varphi  ) }{ E_{\bd{k}}
 + 2 m_s c_{\lambda} ( c_{\lambda} - v ( \bd{k} )  \cos \varphi ) }. 
 \nonumber
 \\
 & &
 \label{eq:gammachex}
 \end{eqnarray}
 where $v ( \bd{k} ) = | \bd{k} | /m_s$, 
the mass $m_s$ is defined via $\rho_s = 1 /( 2 m_s)$, and
the angle-dependent unit vector $\hat{\bd{q}}_{\varphi}$ is defined by
 \begin{eqnarray}
 \hat{\bd{q}}_{\varphi} & = & \left[  \hat{v}_{y}( \bd{k} ) \cos \varphi
 - \hat{v}_{z}( \bd{k} )  \sin \varphi  \right] \bd{e}_y
 \nonumber
 \\
 & + & \left[ \hat{v}_{z}( \bd{k} ) \cos \varphi
 + \hat{ v}_{y}( \bd{k} )  \sin \varphi  \right] \bd{e}_z.
 \label{eq:hatqdef}
 \end{eqnarray}
Here $\hat{{v}}_{\alpha}( \bd{k} ) $ are the components of the unit vector
in the direction of the magnon velocity $\bd{v} ( \bd{k} ) = \nabla_{\bd{k}}
 E_{\bd{k}} = \bd{k} / m_s$.
Although the angular integration in Eq.~(\ref{eq:gammachex}) can be done
analytically, the result is not very transparent so that
we omit it here. 
A graph of the Cherenkov contribution to the
high-temperature damping rate in the exchange regime
is shown as the thin dashed line in Fig.~\ref{fig:damp}.
Next, consider the contribution (\ref{eq:dampconf}) of the confluent scattering
process to the damping rate in the exchange regime.
Setting again $E_{\bd{k}} \approx h + \rho_s {\bd{k}}^2$ and
carrying out the integration over $ | \bd{q} |$ we obtain
 \begin{eqnarray}
 \gamma_2^{\rm con} ( \bd{k} ) & = & \frac{ T E_{\bd{k}}}{4}
 \frac{a^2}{m} \sum_{\lambda} \int_0^{2 \pi} \frac{ d \varphi}{2 \pi}
 \frac{ V_{\lambda}^2 ( \hat{\bd{q}}_{\varphi} ) }{ c_{\lambda}^3 }
 \nonumber
 \\
 & & \times
 \frac{ \Theta ( c_{\lambda} + v ( \bd{k} ) \cos \varphi - \sqrt{ \frac{
 4 E_{\bd{k}}}{m_s}} ) }{
 \sqrt{  (c_{\lambda} + v ( \bd{k} ) \cos \varphi)^2 -  \frac{4E_{\bd{k}}}{m_s}}}
 \nonumber
 \\
  & & \times \left[
 \frac{ c_{\lambda} q_+}{ c_{\lambda} q_+ - E_{\bd{k}} }
 +
  \frac{ c_{\lambda} q_-}{ c_{\lambda} q_- - E_{\bd{k}} }
 \right],
 \label{eq:gammaconex}
 \end{eqnarray}
where
 \begin{eqnarray}
 q_{\pm} & = &  m_s ( c_{\lambda} + v ( \bd{k} ) \cos \varphi )
 \nonumber
 \\
 & & \pm  m_s \sqrt{ ( c_{\lambda} + v ( \bd{k} ) \cos \varphi )^2
 - \frac{ 4 E_{\bd{k}}}{m_s }}.
 \end{eqnarray}
The confluent contribution to the
high-temperature damping rate in the exchange regime
is shown graphically as the thin dotted line in Fig.~\ref{fig:damp}.
As one can see, the magnon damping is strongly $\bd{k}$ dependent. 
In particular, it exhibits peaks at the crossing points of magnon and phonon dispersions as well 
as velocities. It also increases by two orders of magnitude between dipolar and exchange regimes, 
which are dominated by confluence and Cherenkov processes respectively. For the magnon lifetime 
$\tau(\bd{k}) = 1/( 2 \pi \gamma(\bd{k}) )$, this implies values of the order of $50 \, {\rm \mu s}$ in 
the dipolar momentum range, while it can be as low as $480 \, {\rm n s}$ for exchange momenta.

\section{Summary and conclusions}
\label{sec:conclusions}

In this work we have  
studied magneto-elastic interactions 
in  experimentally relevant thin films of the magnetic insulator YIG. 
As the dominant sources 
of magneto-elastic interactions are due to relativistic 
effects \cite{Gurevich96} which cannot be taken into account
within  an effective model containing only
spin degrees of freedom, we have used a semi-phenomenological
approach \cite{Abrahams52}, 
which relies on the quantization of a suitable phenomenological expression 
for the magneto-elastic energy.
For the quantized theory we have then  
carefully derived the momentum dependence 
of the magneto-elastic interaction vertices 
within the framework of the conventional $1/S$-expansion 
for ordered quantum spin systems. 
Using these vertices, 
we have explicitly calculated the leading contributions to
the hybridisation between magnon and phonon modes, 
as well as the damping of the magnons due to spin-lattice coupling. 
The hybridisation has been shown to give rise to a characteristic minimum 
in the spin dynamic structure factor at the crossing point of magnon
and transversal phonon dispersions, where 
the spectral weight is transferred from the magnons to
the transverse phonon mode.
The position of this minimum quantitatively agrees with the recent experimental 
observation of the magneto-elastic mode~\cite{AnnualReport12}. 

The damping at room temperature has been shown to be strongly momentum 
dependent.
In the long-wavelength dipolar regime it is rather flat and 
almost exclusively driven by confluent magnon-phonon scattering processes
where two magnons decay into a phonon or vice versa.
In this regime, we have also presented new experimental results
for the magnon damping obtained by 
wave-vector-resolved Brillouin light scattering spectroscopy.
The fact that the experimental results for the magnon damping are 
roughly three orders of magnitude larger
than our theoretical results indicate that in the dipolar regime
magnon-phonon interactions are not the dominant source of
magnon damping in our samples at room temperature.
We suspect that in this regime the magnon damping is dominated
by elastic scattering of magnons from impurities.
On the other hand, in the short-wavelength exchange regime
the damping is due to magnon-phonon scattering processes of the
Cherenkov type and is
two orders of magnitude
larger than in the dipolar regime.
The damping rate exhibits  pronounced peaks 
at the crossing points of magnon and phonon dispersions and velocities.
This agrees very well with the conclusions of the experiment~[\onlinecite{Agrawal13}], where the authors
suggested that the spin-lattice relaxation in the dipolar regime should be much slower
than in the exchange regime
in order to reconcile their results with 
earlier work on the spin Seebeck effect.

The present work can be extended in two directions:
on the theoretical side, it would be useful
to have quanitatively accurate calculations
of the magnon damping due to magnon-impurity and magnon-magnon interactions
in the dipolar regime; we expect that this can provide
a better 
explanation for our experimental results shown in Fig.~\ref{Figure_1},
which is three orders of magnitude larger than the damping
due to magnon-phonon interactions in this regime.
Note, however, that recently Chernyshev \cite{Chernyshev12}
has considered spontaneous magnon decays 
of the $\bd{k} =0$ magnon in YIG due to magnon-magnon interactions
in high magnetic fields.
On the experimental side, it would be useful to measure magnon damping
in the exchange regime and compare the data with
our theoretical prediction shown in Fig.~\ref{fig:damp}.

\section*{ACKNOWLEDGMENTS} 
Financial support by the DFG via
SFB/TRR49 is gratefully acknowledged.

\begin{appendix}
\renewcommand{\theequation}{A\arabic{equation}}

\section*{APPENDIX:  MAGNETO-ELASTIC MODES FROM EQUATIONS OF MOTION}
\setcounter{equation}{0}
\renewcommand{\theequation}{A\arabic{equation}}

In this appendix we show how to obtain
the energy dispersions of the magneto-elastic modes 
from the linearized equations of motion of the coupled
magnon-phonon system.
Although our derivation using the effective magnon action
presented in Sec.~\ref{sec:hybrid}
is simpler,  the derivation in this appendix 
is more in the spirit of previous work \cite{Gurevich96} using
classical equations of motion.

To obtain the energy dispersions of the eigenmodes
we write down the Heisenberg equations of motion of both
the Holstein-Primakoff bosons and the phonon operators. Within 
the linear approximation we have
 \begin{subequations}
\begin{eqnarray}
 i \dot{b}_{\bd{k}} & = & A_{\bd{k}} b_{\bd{k}} + B_{\bd{k}} 
 b^{\dagger}_{- \bd{k}} + \bd{\Gamma}^{\ast}_{\bd{k}} \cdot
 \bd{X}_{\bd{k}},
 \label{eq:eomb}
 \\
 -i \dot{b}^ {\dagger}_{-\bd{k}} & = & A_{\bd{k}} b^{\dagger}_{-\bd{k}} 
 + B_{\bd{k}} 
 b_{ \bd{k}} + \bd{\Gamma}_{ - \bd{k}} \cdot
 \bd{X}_{\bd{k}},
 \label{eq:eombb}
 \end{eqnarray}
 \end{subequations}
 \begin{subequations}
\begin{eqnarray}
 \dot{X}_{\bd{k} \lambda} & = & \frac{ P_{\bd{k} \lambda}}{m},
 \\
  \dot{P}_{\bd{k} \lambda} & = & - m\omega^2_{\bd{k} \lambda}
  X_{\bd{k} \lambda} - \Gamma_{ \bd{k} \lambda} b_{\bd{k} }
 - \Gamma^{\ast}_{- \bd{k} \lambda} b^{\dagger}_{- \bd{k}}.
 \end{eqnarray}
 \end{subequations}
It is useful to express the equations of motion
for the Holstein-Primakoff bosons
in terms of the transverse spin components,
which to leading order in the $1/S$-expansion can be identified with
 \begin{eqnarray}
 S^x_{\bd{k}} & = & \frac{\sqrt{2S}}{2} \left( b_{\bd{k}} 
 + b^{\dagger}_{- \bd{k}}  \right),
 \\
 S^y_{\bd{k}} & = & \frac{\sqrt{2S}}{2i} \left( b_{\bd{k}} 
 - b^{\dagger}_{- \bd{k}}  \right).
  \end{eqnarray}
Then Eqs.~(\ref{eq:eomb}) and (\ref{eq:eombb}) can be written as
 \begin{subequations}
 \begin{eqnarray}
 \dot{S}^x_{\bd{k}} & = & ( A_{\bd{k}} - B_{\bd{k}} ) S^y_{\bd{k}}
 + i B_{\bot} \bd{k}_{yz} \cdot \bd{X}_{\bd{k}},
 \label{eq:Sx}
 \\
 \dot{S}^y_{\bd{k}} & = & - ( A_{\bd{k}} + B_{\bd{k}} ) S^x_{\bd{k}}
 - i B_{\bot} \bd{k}_{xz} \cdot \bd{X}_{\bd{k}},
 \label{eq:Sy}
 \end{eqnarray}
 \end{subequations}
while the phonon momenta satisfy
 \begin{eqnarray}
  \dot{P}_{\bd{k} \lambda} 
 & = & - m\omega^2_{\bd{k} \lambda}
  X_{\bd{k} \lambda} + i \frac{ B_{\bot} }{ S }
 {\bd{e}}^{\ast}_{\bd{k} \lambda} \cdot
 \left( \bd{k}_{xz} S^x_{\bd{k}} +  \bd{k}_{yz} S^y_{\bd{k}} \right),
 \nonumber
 \\
 & &
\end{eqnarray} 
implying
 \begin{eqnarray}
 \ddot{ X}_{\bd{k} \lambda} +  \omega^2_{\bd{k} \lambda}
  X_{\bd{k} \lambda} =  i \frac{B_{\bot}}{m S}
 {\bd{e}}^{\ast}_{\bd{k} \lambda} \cdot
 \left( \bd{k}_{xz} S^x_{\bd{k}} +  \bd{k}_{yz} S^y_{\bd{k}} \right).
 \end{eqnarray}
If we ignore the magnon-phonon coupling, we obtain from 
Eqs.~(\ref{eq:Sx},\ref{eq:Sy}),
 \begin{equation}
 \ddot{S}^{\alpha}_{\bd{k}} + E_{\bd{k}}^2 {S}^{\alpha}_{\bd{k}} =0,
 \; \; \;  \alpha = x,y,
 \end{equation}
where the magnon dispersion in the absence of phonons is
given in Eq.~(\ref{eq:Ekdef}).
With finite magnon-phonon hybridization we obtain the 
energies of the magneto-elastic modes from the roots of the
secular determinant of the above equations of motion.
For simplicity, let us assume that the energy of only 
one particular phonon mode $\omega_{\bd{k} \lambda}$
is close to $E_{\bd{k}}$. To calculate the
energy of the magneto-elastic mode close to the crossing point,
it is then sufficient to approximate $\bd{X}_{\bd{k}} \approx
 X_{\bd{k} \lambda} \bd{e}_{\bd{k} \lambda}$ in the above equations of motion.
Then we obtain the energies of the magneto elastic modes
from the roots of the following quartic secular equation,
 \begin{eqnarray}
 & & ( \omega^2 - \omega_{\bd{k} \lambda}^2 ) ( \omega^2 - E_{\bd{k}}^2 )
 \nonumber
 \\
 & = & \frac{ B_{\bot}^2}{m S}
 \Bigl\{ (A_{\bd{k}} + B_{\bd{k}} ) 
 | \bd{k}_{yz} \cdot \bd{e}_{\bd{k} \lambda} |^2
 +
 (A_{\bd{k}} - B_{\bd{k}} ) 
 | \bd{k}_{xz} \cdot \bd{e}_{\bd{k} \lambda} |^2 
 \nonumber
 \\
 & & \hspace{7mm} + 2 \omega  {\rm Im} \left[
 ( \bd{k}_{xz} \cdot \bd{e}_{\bd{k} \lambda} )
 ( \bd{k}_{yz} \cdot \bd{e}^{\ast}_{\bd{k} \lambda} )
 \right]
 \Bigr\}.
 \label{eq:secular}
 \end{eqnarray}
For the phonons in a thin YIG stripe the
basis vectors $\bd{e}_{\bd{k} \lambda}$ can always be chosen such
that the last term in Eq.~(\ref{eq:secular}) vanishes,
so that the secular equation is bi-quadratic and can be
explicitly solved. The square of the
energies of the magneto-elastic modes in a thin YIG stripe are therefore
 \begin{equation}
 ( \Omega^{\pm}_{\bd{k} \lambda } )^2 =
 \frac{ \omega^2_{\bd{k} \lambda} + E^2_{\bd{k}}}{2}
 \pm \sqrt{  \frac{ (\omega^2_{\bd{k} \lambda} - E^2_{\bd{k}})^2 }{4}
 + \Delta_{\bd{k} \lambda}^4 },
 \label{eq:omegaeom}
 \end{equation}
where
 \begin{eqnarray}
  \Delta_{\bd{k} \lambda}^4 & = &
 B_{\bot}^2
 \Bigl[ (A_{\bd{k}} + B_{\bd{k}} ) 
 \frac{ | \bd{k}_{yz} \cdot \bd{e}_{\bd{k} \lambda} |^2}{m S}
 \nonumber
 \\ 
 &  &  \hspace{4mm} +(A_{\bd{k}} - B_{\bd{k}} ) 
 \frac{| \bd{k}_{xz} \cdot \bd{e}_{\bd{k} \lambda} |^2 }{m S}
 \Bigr].
 \end{eqnarray}
If we approximate $B_{\bd{k}} \approx 0$ (which corresponds to
neglecting quantum fluctuations generated by the
dipolar interaction) Eq.~(\ref{eq:omegaeom}) agrees
with the result obtained via the 
classical equations of motion  \cite{Gurevich96}.

\end{appendix}

\end{document}